\documentclass[onecolumn]{aastex6}

\usepackage{amsmath}    % mathematical notation
\usepackage{mathtools}  % paired delimiters
\usepackage{tensor}     % tensor notation
\usepackage{hyperref}   % links
\hypersetup{colorlinks=true,linkcolor=black,citecolor=black,filecolor=black,urlcolor=black}

\makeatletter
\newcommand{\extref}[1]{\textup{\tagform@{#1}}}
\makeatother

\newcommand{\ee}{\mathrm{e}}
\newcommand{\dd}{\mathrm{d}}
\newcommand{\dt}{\dd t}
\newcommand{\dr}{\dd r}
\newcommand{\dth}{\dd\theta}
\newcommand{\dph}{\dd\phi}
\newcommand{\drr}{\dd R}
\newcommand{\dvph}{\dd\varphi}
\newcommand{\dz}{\dd z}
\newcommand{\cyl}{_\mathrm{cyl}}
\newcommand{\sph}{_\mathrm{sph}}
\newcommand{\midp}{_\mathrm{mid}}
\DeclarePairedDelimiter{\paren}{\lparen}{\rparen}
\DeclarePairedDelimiter{\abs}{\lvert}{\rvert}
\DeclarePairedDelimiter{\ave}{\langle}{\rangle}

\newcommand{\rhor}{r_\mathrm{hor}}

\newcommand{\redge}{r_\mathrm{edge}}
\newcommand{\rmin}{r_\mathrm{min}}
\newcommand{\rmax}{r_\mathrm{max}}
\newcommand{\thetamin}{\theta_\mathrm{min}}
\newcommand{\thetamax}{\theta_\mathrm{max}}
\newcommand{\pgas}{p_\mathrm{gas}}
\newcommand{\pmag}{p_\mathrm{mag}}
\newcommand{\rhomin}{\rho_\mathrm{min}}
\newcommand{\pgasmin}{p_\mathrm{gas,min}}
\newcommand{\tmin}{t_\mathrm{min}}
\newcommand{\tmax}{t_\mathrm{max}}
\newcommand{\vaz}{v_{\mathrm{A},z}}
\newcommand{\gammaaz}{\gamma_{\mathrm{A},z}}
\newcommand{\lambdac}{\lambda_\mathrm{crit}}
\newcommand{\gammac}{\gamma_\mathrm{crit}}
\newcommand{\fcrit}{f_\mathrm{crit}}
\newcommand{\lambdam}{\lambda_\mathrm{max}}
\newcommand{\fmax}{f_\mathrm{max}}

\newcommand{\msun}{M_\odot}
\newcommand{\mpc}{\mathrm{Mpc}}
\newcommand{\mmas}{\mathrm{\mu as}}
\newcommand{\ghz}{\mathrm{GHz}}
\newcommand{\jy}{\mathrm{Jy}}

\newcommand{\athena}{\texttt{Athena++}}
\newcommand{\ibothros}{\texttt{ibothros}}

\shorttitle{MAD Resolution}
\shortauthors{C.~J.~White, J.~M.~Stone, E.~Quataert}

\begin{document}

% Title and author information
\title{A Resolution Study of Magnetically Arrested Disks}
\author{Christopher~J.~White\altaffilmark{1}, James~M.~Stone\altaffilmark{2}, and Eliot~Quataert\altaffilmark{3}}
\altaffiltext{1}{Kavli Institute for Theoretical Physics, University of California Santa Barbara, Kohn Hall, Santa Barbara, CA 93107, USA}
\altaffiltext{2}{Department of Astrophysical Sciences, Princeton University, 4 Ivy Lane, Princeton, NJ 08544, USA}
\altaffiltext{3}{Department of Astronomy, University of California Berkeley, 501 Campbell Hall, Berkeley, CA 94720, USA}

% Abstract
\begin{abstract}
We investigate numerical convergence in simulations of magnetically arrested disks around spinning black holes. Using the general-relativistic magnetohydrodynamics code \athena, we study the same system at four resolutions (up to an effective $512\times256\times512$ cells) and with two different spatial reconstruction algorithms. The accretion rate and general large-scale structure of the flow agree across the simulations. This includes the amount of magnetic flux accumulated in the saturated state and the ensuing suppression of the magnetorotational instability from the strong field. The energy of the jet and the efficiency with which spin energy is extracted via the Blandford--Znajek process also show convergence. However the spatial structure of the jet shows variation across the set of grids employed, as do the Lorentz factors. Small-scale features of the turbulence, as measured by correlation lengths, are not fully converged. Despite convergence of a number of aspects of the flow, modeling of synchrotron emission shows that variability is not converged and decreases with increasing resolution even at our highest resolutions.
\end{abstract}

% Keywords

% Introduction
\section{Introduction}
\label{sec:introduction}

The notion that accumulation of ordered magnetic field could qualitatively change the nature of an accretion flow was first investigated by \citet{BisnovatyiKogan1974}. Numerical simulations by \citet{Narayan2003} confirmed that poloidal flux could lead to an efficient conversion of gravitational energy into mechanical energy, even around a non-spinning black hole. These simulations showed a magnetically arrested disk (MAD), in which accretion was hampered by a strong magnetic barrier.

Simulations around spinning black holes followed \citep{Hawley2004,Tchekhovskoy2011,McKinney2012}, adding in some cases optically thin cooling \citep{Avara2016} or radiative transfer \citep{McKinney2015,MoralesTeixeira2018}. In some of these simulations, the outflowing energy was found to exceed the amount infalling. This can be attributed to the Blandford--Znajek process \citep{Blandford1977} at play, electromagnetically extracting energy from the spin of the black hole and using it to launch a jet. At the same time, magnetic field threading a disk is expected to launch a wind via the Blandford--Payne process \citep{Blandford1982}.

\Citet{McKinney2012} ran a number of models in which they varied physical parameters such as black hole spin and magnetic field geometry to explore the MAD parameter space. Here we address the orthogonal issue of to what extent numerical results depend on numerical parameters. That is, how well do simulations of this nature capture all the relevant physics? In particular, four concerns we seek to address are:

\begin{enumerate}

  \item In axisymmetry, MAD leads to a completely arrested inflow that is broken by occasional bursts of accretion, while fully 3D simulations show interchanges of dense, less-magnetized fluid with magnetically buoyant bubbles \citep{Igumenshchev2008}. The dynamics of a heavier fluid supported against gravity by a lighter fluid lend themselves to the magnetic Rayleigh--Taylor instability \citep[RTI][]{Kruskal1954}, only recently studied analytically in this context \citep{Contopoulos2016}. Without sufficient resolution, one might not fully capture this nonaxisymmetric instability, and this can in turn affect the reported accretion of matter and magnetic flux.

  \item One defining characteristic of the MAD state is suppression of the magnetorotational instability \citep[MRI,][]{Balbus1991}. However, merely knowing that the MRI is suppressed is not sufficient to determine the strength of turbulence in these flows, as mechanisms such as the aforementioned RTI can also induce turbulence, as for instance is found by \citet{Marshall2018}. What resolutions sufficiently resolve turbulence?

  \item However much or little the statistical nature of a turbulent disk may change with resolution, how much does this affect the smoother, highly magnetized, relativistic jet? Jets are an extremely important consequence of MAD conditions, especially given their observability. What resolutions are needed to be confident in simulations' measured jet properties?

  \item What impact does grid resolution have on directly observable quantities?

\end{enumerate}

In order to address these concerns, we run a single MAD scenario around a rotating black hole, varying the resolution but not the physical parameters. We go up to an effective resolution of $173$ cells per decade in radius, $256$ cells in polar angle, and $512$ cells in azimuthal angle. Our simulations are performed with the GRMHD capabilities of \athena{} \citep{White2016}, which utilizes staggered-mesh constrained transport and transmissive rather than reflecting polar axis boundary conditions. We employ the HLLE Riemann solver \citep{Einfeldt1988} commonly used in the GRMHD community. While this solver is not the least diffusive available, we have found that having even less diffusion adversely affects the robustness of these highly magnetized simulations around rapidly spinning black holes. The code can use either a piecewise linear method \citep[PLM,][]{VanLeer1974} or the piecewise parabolic method \citep[PPM,][]{Colella1984} for spatial reconstruction. In multiple dimensions both methods formally converge at second order in spatial resolution, though the latter often has significantly smaller errors. We run simulations with both methods, since the effect of using PPM over PLM is to better capture spatial variability, just as would occur with increasing resolution.

In \S\ref{sec:simulations} we provide details about the setup and running of the simulations. We then investigate the effect of resolution on the disk itself (\S\ref{sec:disk}), on the jetted outflow (\S\ref{sec:jet}), and on observable electromagnetic signatures (\S\ref{sec:light_curves}). Discussion and conclusions follow in \S\ref{sec:discussion} and \S\ref{sec:conclusion}.

% Simulations
\section{Simulations}
\label{sec:simulations}

All of our simulations are done in the same spherical Kerr-Schild coordinates $(t,r,\theta,\phi)$ as defined in \citet{Gammie2003}, where we will drop $c$, $G$, and the black hole mass $M$ from all expressions. The coordinates are regular at the outer horizon $\rhor = 1 + \sqrt{1-a^2}$ and have determinant given by $\sqrt{-g} = (r^2 + a^2 \cos^2\!\theta) \sin\theta$. We set the spin to be $a = 0.98$. In some cases we will express quantities in the cylindrical version of these coordinates, obtained with the standard spherical-to-cylindrical transformations. That is, we have $(t,R,\phi,z)$, where $R = r \sin\theta$ and $z = r \cos\theta$. The metric determinant in these coordinates is given by $\sqrt{-g\cyl} = \sqrt{-g} / r$. In what follows we will denote fluid-frame rest-mass density by $\rho$; fluid-frame gas pressure by $\pgas$; $4$-velocity components by $u^\mu$; normal-frame Lorentz factor by $\gamma$; fluid-frame magnetic pressure by $\pmag \equiv b_\mu b^\mu / 2$, where $b^\mu \equiv u_\nu (*F)^{\nu\mu}$ and $F$ is the electromagnetic field tensor; magnetic field components by $B^i \equiv (*F)^{i0}$; and the plasma magnetization parameters by $\beta^{-1} \equiv \pmag/\pgas$ and $\sigma \equiv 2\pmag/\rho$.

We begin with a torus of material in hydrodynamical equilibrium, following the prescription of \citet{Fishbone1976}. We set the inner edge of the disk to be at $r = 16.45$, with the pressure maximum at $r = 34$. The fluid has a fixed adiabatic index $\Gamma = 13/9$. This value is chosen as a compromise between relativistically hot ($\Gamma = 4/3$) electrons and relativistically cold ($\Gamma = 5/3$) ions, as is done for example by \citet{Ryan2018} when more carefully considering the two-temperature nature of the accreting plasma. We add an initial magnetic field similar to that of \citet{Tchekhovskoy2011}. That is, the initial field is purely poloidal, circulating about a point well outside the pressure maximum. Specifically, we take the vector potential to have only an azimuthal component
\begin{equation}
  A_\phi \propto \paren[\big]{\max(\pgas - 10^{-8}, 0)}^{1/2} r^2 \sin\theta \sin\paren[\big]{\pi L(r; 17, 500)} \sin\paren[\big]{\pi L(\theta; \pi/6, 5\pi/6)},
\end{equation}
where $L$ is the linear ramp function
\begin{equation}
  L(x; x_\mathrm{min}, x_\mathrm{max}) \equiv
  \begin{cases}
    0, & x \leq x_\mathrm{min}; \\
    \dfrac{x-x_\mathrm{min}}{x_\mathrm{max}-x_\mathrm{min}}, & x_\mathrm{min} < x < x_\mathrm{max}; \\
    1, & x \geq x_\mathrm{max}.
  \end{cases}
\end{equation}
The field is normalized to have the density-weighted average of $\beta^{-1}$ of the torus be $0.01$.

\begin{figure}
  \centering
  \includegraphics{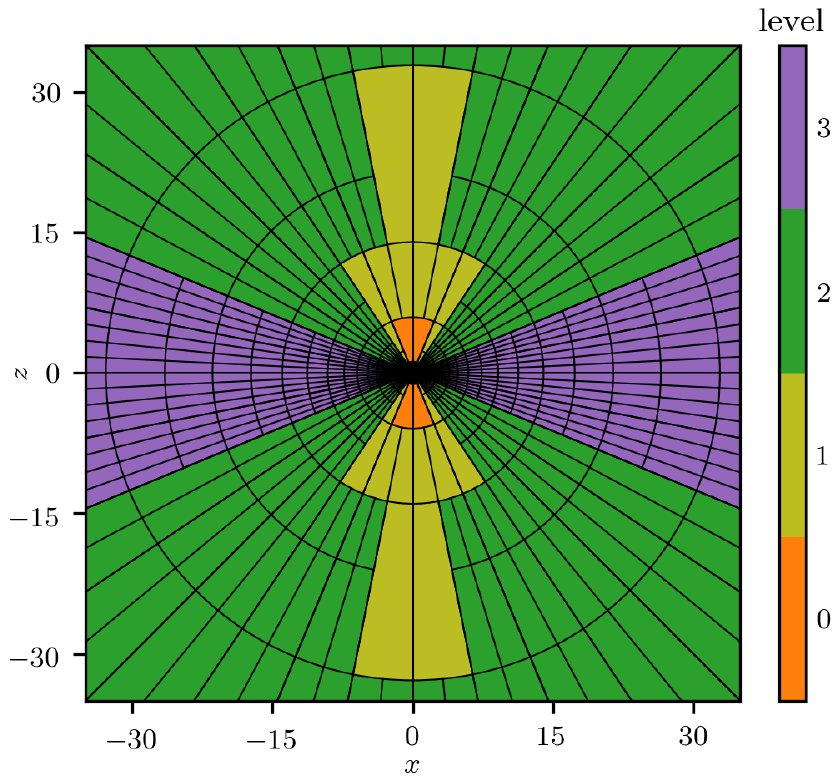}
  \caption{Inner portion of the fully refined (level~3) grid used in these simulations. Each box represents a block of $16$ cells in radius and $4$ cells in polar angle. When we refer to a simulation on a level~$n$ grid, the grid will resemble the above picture, but with levels $n+1$ and higher coarsened to level $n$. \label{fig:grid}}
\end{figure}

Our root grid (refinement level 0) has $64\times32\times64$ cells in $r$, $\theta$, and $\phi$. In radius it extends from $r \approx 1.078$ to $r = 1000$ with logarithmic spacing, with the innermost cell entirely inside the horizon. The grid is uniform in $\theta$ and $\phi$. We then add successive layers of static mesh refinement to the domain as follows. Each additional level refines all cells in its domain by a factor of $2$ in each dimension. Level~1 refinement covers $\pi/8 < \theta < 7\pi/8$ at all radii and all polar angles for $r \gtrsim 5.948$. Level~2 covers $3\pi/16 < \theta < 13\pi/16$ at all radii, $\pi/16 < \theta < 15\pi/16$ for $r \gtrsim 13.97$, and all polar angles for $r \gtrsim 32.83$. Level~3 covers $3\pi/8 < \theta < 5\pi/8$ for $r \lesssim 118.2$. The grid is illustrated in Figure~\ref{fig:grid}, where the grid lines are drawn every $16$ cells in $r$ and every $4$ cells in $\theta$.

With this grid, the timestep at all refinement levels is limited by the $\phi$-width of the (never-refined) cells at the base of the polar axis. The root grid has a radial resolution of approximately $22$ cells per decade, and each successive refinement level doubles this, up to $173$ cells per decade at the highest resolution. At the highest resolution there are $512$ cells in azimuth, and the polar resolution in the bulk of the disk is the same as an effective $256$ equally spaced cells. These values are detailed in Table~\ref{tab:refinement}.

\floattable
\begin{deluxetable}{cCCCCCC}
  \tablecaption{Description of static mesh refinement regions. \label{tab:refinement}}
  \tablewidth{0pt}
  \tablehead{\colhead{Level} & \colhead{$r$-range\tablenotemark{a}} & \colhead{$N_{r,\mathrm{eff}}$\tablenotemark{b}} & \colhead{cells/decade} & \colhead{$\theta$-range\tablenotemark{c}} & \colhead{$N_{\theta,\mathrm{eff}}$\tablenotemark{b}} & \colhead{$N_\phi$}}
  \decimals
  \startdata
  0 & 1.078\text{--}1000\phd\phn & \phn64 & \phn22 & 0\text{--}\pi                & \phn32 & \phn64 \\
  1 & 1.078\text{--}1000\phd\phn & 128    & \phn43 & \phn\pi/8\text{--}7\pi/8     & \phn64 & 128    \\
  2 & 1.078\text{--}1000\phd\phn & 256    & \phn86 & \phn3\pi/16\text{--}13\pi/16 & 128    & 256    \\
  3 & 1.078\text{--}\phd118.2    & 512    & 173    & 3\pi/8\text{--}5\pi/8        & 256    & 512    \\
  \enddata
  \tablenotetext{a}{Evaluated at the midplane. Range may be less at higher latitudes.}
  \tablenotetext{b}{Number of cells needed to extend resolution within range to entire domain.}
  \tablenotetext{c}{Evaluated at small radii. Range may differ at large radii.}
\end{deluxetable}

For comparison, the grid employed in \citet{Tchekhovskoy2011} and in model A0.99N100 of \citet{McKinney2012} (the only one with a spin approximately the same as ours) has $128$ cells in polar angle, $64$ cells in azimuth (doubled to $128$ midway through the simulation) and fewer than $95$ cells per decade in radius. The high-resolution runs A0.94BfN40, A-0.94BfN40HR, and A0.94BtN10HR of \citeauthor{McKinney2012}\ all have $87\text{--}88$ cells per radial decade in the inner regions, $425$ effective polar cells based on the midplane resolution at $r = 10$, and $256$ cells in azimuth. A more recent study on MAD accretion, \citet{Avara2016}, has an effective resolution of approximately $600$ cells in $\theta$ (used to resolve a thin disk) and $208$ cells in $\phi$, while the tilted jets in \citet{Liska2018} are simulated with effective resolutions up to $576\times960$ in angle. Except for this last case, our highest resolution typically has $2$ or more times as many total cells per dimension as these previous calculations.

We define the radial width $W_r$ of a cell to be the integral of $\sqrt{g_{rr}}$ over the line of constant $\theta$ and $\phi$ passing through the cell midpoint from one constant-$r$ face to another, and likewise for polar and azimuthal widths. (These reduce to the expected $\Delta r$, $r \Delta\theta$, and $r \sin\theta \Delta\phi$ in the Newtonian limit.) Then in the midplane our grid has $W_\theta/W_r$ varying from approximately $0.6522$ at the inner boundary to $0.9194$ at the outer boundary, while $W_\phi/W_r$ varies from $1.138$ to $0.9194$. Having aspect ratios near unity across all our grids is important, as refining only one or two dimensions will have little effect on reducing errors once those errors become dominated by the coarser dimensions.

A number of limits are imposed on the variables for numerical stability. The density and gas pressure are kept above the floors
\begin{subequations} \begin{align}
  \rhomin & = \max\paren[\big]{10^{-4} (r/M)^{-3/2}, 10^{-8}}, \\
  \pgasmin & = \max\paren[\big]{10^{-6} (r/M)^{-5/2}, 10^{-10}}.
\end{align} \end{subequations}
In addition, we enforce $\beta^{-1} < 100$, $\sigma < 100$, and $\gamma < 50$. Whenever these limits are imposed, the magnetic field is left unaltered. For comparison, \citet{Tchekhovskoy2011} enforce either $\beta^{-1} < 7500$ and $\sigma < 500$ or else $\beta^{-1} < 750$ and $\sigma < 50$ (stating that the choice makes little difference), while \citet{McKinney2012} have $\beta^{-1} < 1500$ and $\sigma < 50$. More recent work by \citet{Liska2018}, which includes strong fields launching jets, uses the limits $\beta^{-1} < 225$ and $\sigma < 50$, while also enforcing $\rho > 10^{-6} (r/M)^{-2}$ and $\pgas > 2/3 \cdot 10^{-7} (r/M)^{-10/3}$.

We start with a single simulation with level~3 refinement and PPM reconstruction, running until time $10{,}000$. This checkpoint is then coarsened to the other grids, and all simulations are then run from $t = 10{,}000$ to $t = 20{,}000$, using both PLM and PPM. When averaging results in time, we choose $15{,}000 \leq t \leq 20{,}000$ in all cases.

To run for a simulation time of $10{,}000$, the level~3 PPM simulation took $27{,}600$ node-hours on $125$ Intel Knights Landing (Xeon Phi 7250) nodes, with $68$ cores per node. The level~2 PPM simulation took $3990$ node-hours on $37$ Intel Skylake (Xeon Platinum 8160) nodes, with $48$ cores per node.

% Disk properties
\section{Disk Properties}
\label{sec:disk}

% Disk properties: Mass accretion rate
\subsection{Mass Accretion Rate}
\label{sec:disk:mass}

\begin{figure*}
  \centering
  \includegraphics{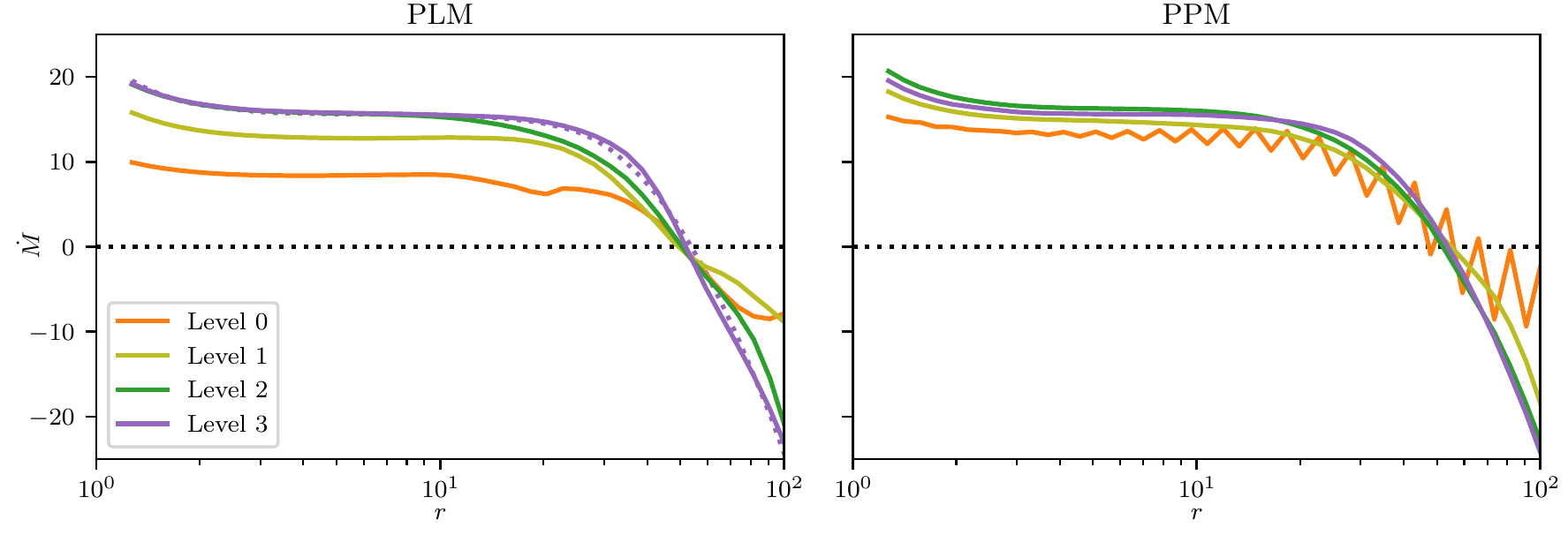}
  \caption{Time-averaged mass accretion rate as a function of radius. The level~3 PPM curve is duplicated on the left with a dotted line for comparison. The level~2 and level~3 simulations, using both PLM and PPM, agree on the steady-state value. The rate is underestimated with insufficient resolution. Oscillatory behavior occurs at very low resolution with PPM, but in no other cases. \label{fig:mdot_radius}}
\end{figure*}

We define the mass accretion rate at a fixed time through a sphere $S(r)$ of fixed radius $r$ to be
\begin{equation} \label{eq:mdot}
  \dot{M}(r) \equiv -\int\limits_{S(r)} \rho u^r \sqrt{-g} \, \dth\,\dph.
\end{equation}
Positive values correspond to inflow. Figure~\ref{fig:mdot_radius} shows the run of $\dot{M}$ with $r$, averaged in time, for our simulations. In steady state, we expect $\dot{M}$ to be independent of radius, and indeed we do see an extended plateau. There is a small rise inward at very small radii ($r \lesssim 2$). This is an artifact of numerical floors, and it measures the small amount of mass added by the floors at these radii. Other simulations using different codes see the same effect; see for example Figure~4 of \citet{MoralesTeixeira2018}. The four level~2 and level~3 simulations agree on the plateau height. At sufficiently low resolutions, the accretion rate is underestimated. Moreover, there is a cell-to-cell oscillatory numerical artifact that occurs at very low resolutions when using PPM, almost certainly due to the slope limiter itself. There is no sign of this effect at higher resolutions.

\begin{figure*}
  \centering
  \includegraphics{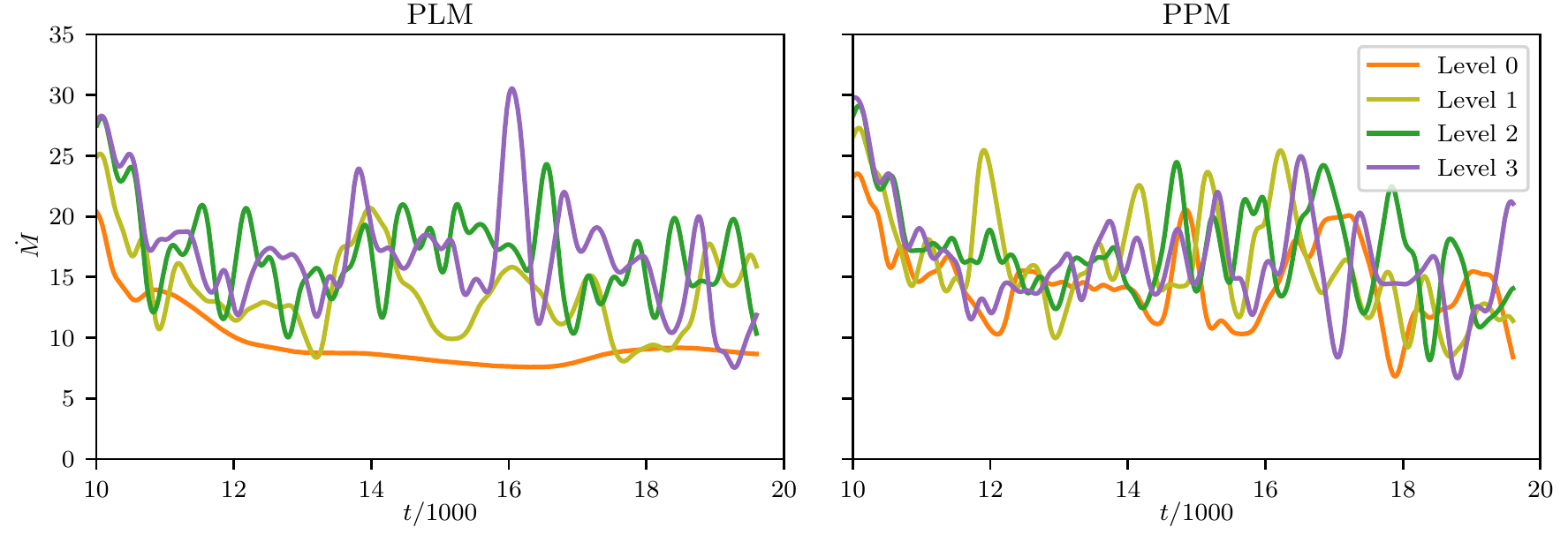}
  \caption{Mass accretion rate as a function of time, smoothed with a Gaussian filter of width $\Delta t = 100$. At sufficiently low resolution with only PLM reconstruction, turbulence does not fully develop. In the other cases, the flow is highly variable. \label{fig:mdot_time}}
\end{figure*}

When fixing a single radius at which to measure $\dot{M}$, we choose $r = 5$ from here on. Using this radius, we show the values of $\dot{M}$ as a function of time in Figure~\ref{fig:mdot_time}. In all cases except the lowest-resolution PLM simulation, the accretion becomes highly variable, experiencing order-unity changes even on the timescale of $\Delta t = 10$ between snapshots. In order to make the figure clearer, we smooth the accretion rate with a Gaussian filter of full-width at half-maximum $\Delta t = 100$. (This smoothing accounts for the lines differing at $t = 10{,}000$, when all the simulations are identical.) The average accretion rates are not changing significantly over time, and there are many oscillations in the time window from $15{,}000$ to $20{,}000$, justifying our time averages. We note that recent work by \citet{Balbus2017} suggests unstable modes near the innermost stable circular orbit (ISCO) can disrupt the inner parts of disks, suggesting caution when searching for and interpreting steady-state values of $\dot{M}$.

% Disk properties: Accumulation of magnetic flux
\subsection{Accumulation of Magnetic Flux}
\label{sec:disk:flux}

\begin{figure*}
  \centering
  \includegraphics{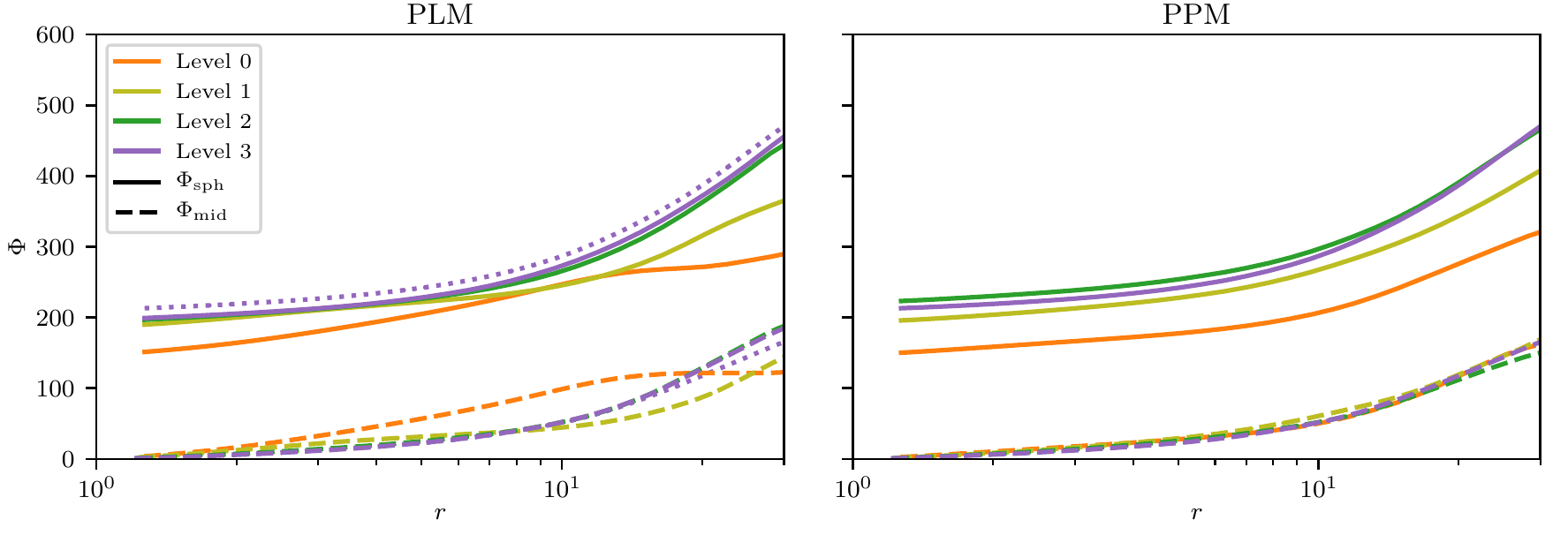}
  \caption{Total magnetic flux $\Phi$ interior to a radius $r$. $\Phi\sph$ measures all flux passing through the sphere, while $\Phi\midp$ only measures flux passing through the midplane. The difference between the solid and dashed lines counts field lines that pass through the sphere of radius $r$ only once, terminating at the horizon rather than continuing through the opposite hemisphere. This difference is what should be constant with radius in steady state. The level~3 PPM curves are duplicated on the left with dotted lines for comparison. \label{fig:flux_radius}}
\end{figure*}

The MAD state is characterized by strong poloidal magnetic flux, which we can measure in a number of ways. Following \citet{Tchekhovskoy2011} and \citet{McKinney2012} we can integrate radial flux through spheres $S(r)$ of different radii $r$ to define
\begin{equation}
  \Phi\sph(r) \equiv \frac{1}{2} \int\limits_{S(r)} \sqrt{4\pi} \abs{B^r} \sqrt{-g} \, \dth\,\dph.
\end{equation}
Our explicit factor of $\sqrt{4\pi}$ is implicit in the units in \citet{Tchekhovskoy2011}. We can also define the midplane vertical flux out to a radius $r$ by either of two equivalent formulas:
\begin{equation}
  \Phi\midp(r) \equiv \int_0^{2\pi}\!\int_{\redge}^r \sqrt{4\pi} B^z \sqrt{-g\cyl} \, \drr\,\dvph \equiv \int_0^{2\pi}\!\int_{\redge}^r \sqrt{4\pi} (-B^\theta) \sqrt{-g} \, \dr\,\dph.
\end{equation}
$\Phi\midp$ counts only flux that passes through the midplane, while $\Phi\sph$ also includes field lines that terminate at the horizon. Figure~\ref{fig:flux_radius} shows both measures of flux, time-averaged, as a function of radius. As $\Phi\sph \gg \Phi\midp$ for the inner part of the disk, most of the magnetic field lines here terminate at the horizon. These profiles agree in the region where steady-state accretion has been attained. Thus even at low resolution we are capturing the large-scale structure of the magnetic field.

\begin{figure*}
  \centering
  \includegraphics{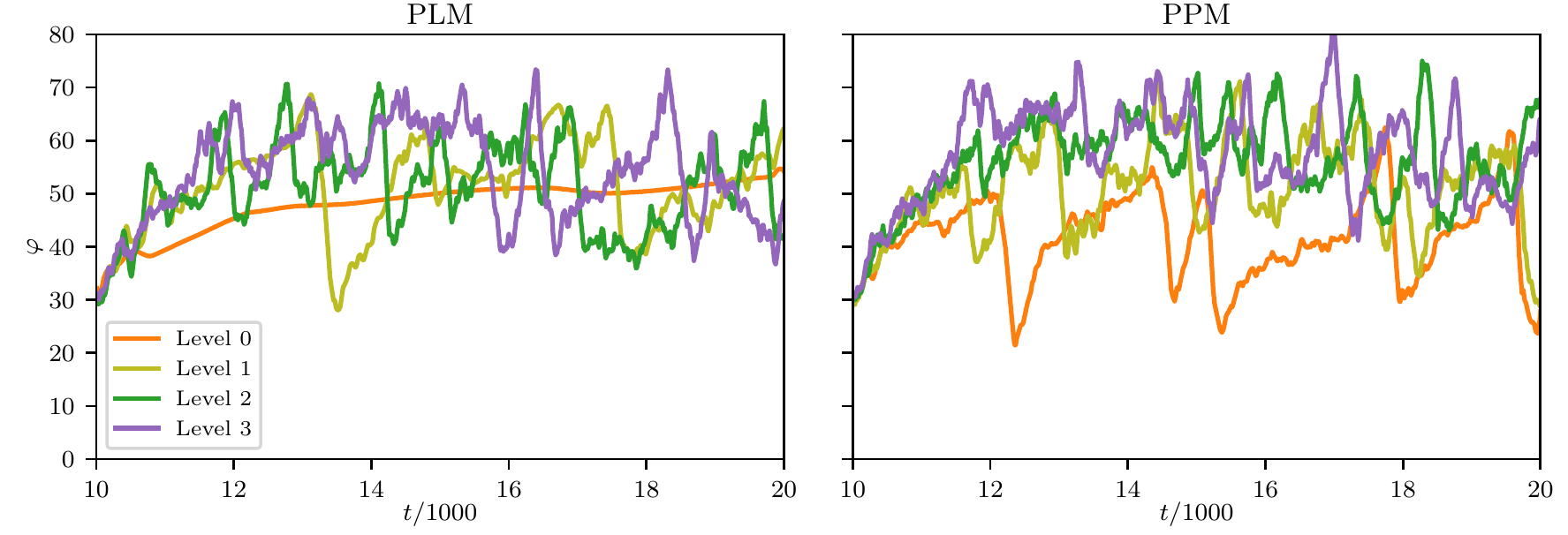}
  \caption{Normalized magnetic flux on the horizon $\varphi$ as a function of time. Most of our simulations are in the expected MAD state. With insufficient resolution, the level and variability are too low (as in level~0 PPM) or there is no variability at all (level~0 PLM), but the solution converges rapidly with increasing resolution. \label{fig:flux_time}}
\end{figure*}

As in \citet{Tchekhovskoy2011}, we normalize the horizon flux based on each simulation's average accretion rate. We choose to use the difference $\Phi\sph - \Phi\midp$, as the value of this quantity at any radius is the flux on the horizon so long as the field is essentially vertical (all field lines are either outside the sphere and midplane annulus, cross the sphere twice and the midplane once, or cross the sphere once and terminate on the horizon). We define
\begin{equation}
  \varphi \equiv \frac{\Phi\sph - \Phi\midp}{\ave{\dot{M}}^{1/2}}.
\end{equation}
Here the fluxes and accretion rate are evaluated at $r = 5$. The time-averaging is done with a Gaussian filter of full-width at half-maximum $100$. Figure~\ref{fig:flux_time} shows the results. The average level and variability in $\varphi$ for most of our cases is similar to that found by \citet{Tchekhovskoy2011}, where the saturated state has $\varphi \approx 47$. (Note there exists another common flux normalization, denoted $\Upsilon$, that is approximately a factor of $6$ times smaller than $\varphi$ at the horizon for our spin.) At the lowest resolution, PPM can still capture some of the variability, though the accumulated flux is too low. At that resolution, PLM gives no time variability, again indicating the numerics in that case are not sufficiently capturing the turbulent nature of the flow.

% Disk properties: The magnetorotational instability
\subsection{The Magnetorotational Instability}
\label{sec:disk:mri}

The crucial qualitative difference between a true MAD state and a disk with just a strong field is the inhibition of the MRI due to the field. In order to investigate if this has indeed occurred, we compare the marginally stable wavelength $\lambdac$ for the vertical MRI to the disk scale height.

The general-relativistic dispersion relation relating Alfv\'en frequency $\gammaaz = k_z \vaz$ to oscillation frequency $\omega$ for a mode is given by \citet{Gammie2004}:
\begin{equation}
  \omega^4 - \omega^2 (2\gammaaz^2 + \kappa^2) + \gammaaz^2 (\gammaaz^2 - s^2) = 0.
\end{equation}
Here $\kappa$ is the epicyclic frequency and we have
\begin{equation}
  s^2 \equiv \frac{3}{r^3} \paren[\bigg]{\frac{1 - 2 r^{-1} + a^2 r^{-2}}{1 - 3 r^{-1} + 2 a r^{-3/2}}}.
\end{equation}
The dispersion relation is quadratic in $\omega^2$, and the critical frequency $\gammac$ corresponds to the value of $\gammaaz$ such that the lesser quadratic root for $\omega^2$ vanishes. This is simply $\gammac^2 = s^2$, as can be easily verified. Now $\gammaaz$ is a frequency as measured by a local observer on a circular orbit. \Citeauthor{Gammie2004}\ gives the prescription for transforming this into a frequency as seen by an observer at infinity. The result is the instability criterion
\begin{equation}
  \lambda > \lambdac \equiv 2\pi \vaz \sqrt{\frac{r^3}{3}} \fcrit,
\end{equation}
where
\begin{equation}
  \fcrit \equiv \frac{1 + a r^{-3/2}}{\paren{1 - 2 r^{-1} + a^2 r^{-2}}^{1/2}}
\end{equation}
is the relativistic correction to the standard formula \citep[cf.\ the nonrelativistic version \extref{108} from][]{Balbus1998}. This correction necessarily approaches unity as $r \to \infty$; it also diverges as $r \to \rhor$, meaning the effect of general relativity is to stabilize the inner disk. Note some sources omit the $\sqrt{3}$ in their definition of $\lambdac$.

For the scale height, we adopt a definition similar to that found in \citet{McKinney2012}, though we vary vertical coordinate $z$ rather than polar coordinate $\theta$. That is, for given $t$, $\phi$, and $R$, we calculate
\begin{equation}
  z\midp \equiv \frac{\int_{z_-}^{z_+} z \rho \sqrt{-g\cyl} \, \dz}{\int_{z_-}^{z_+} \rho \sqrt{-g\cyl} \, \dz},
\end{equation}
where the cutoffs are chosen to be $z_\pm \equiv \pm50$. Then the scale height $H$ at that location is given by
\begin{equation}
  H^2 \equiv \frac{\int_{z_-}^{z_+} (z - z\midp)^2 \rho \sqrt{-g\cyl} \, \dz}{\int_{z_-}^{z_+} \rho \sqrt{-g\cyl} \, \dz}.
\end{equation}

\begin{figure*}
  \centering
  \includegraphics{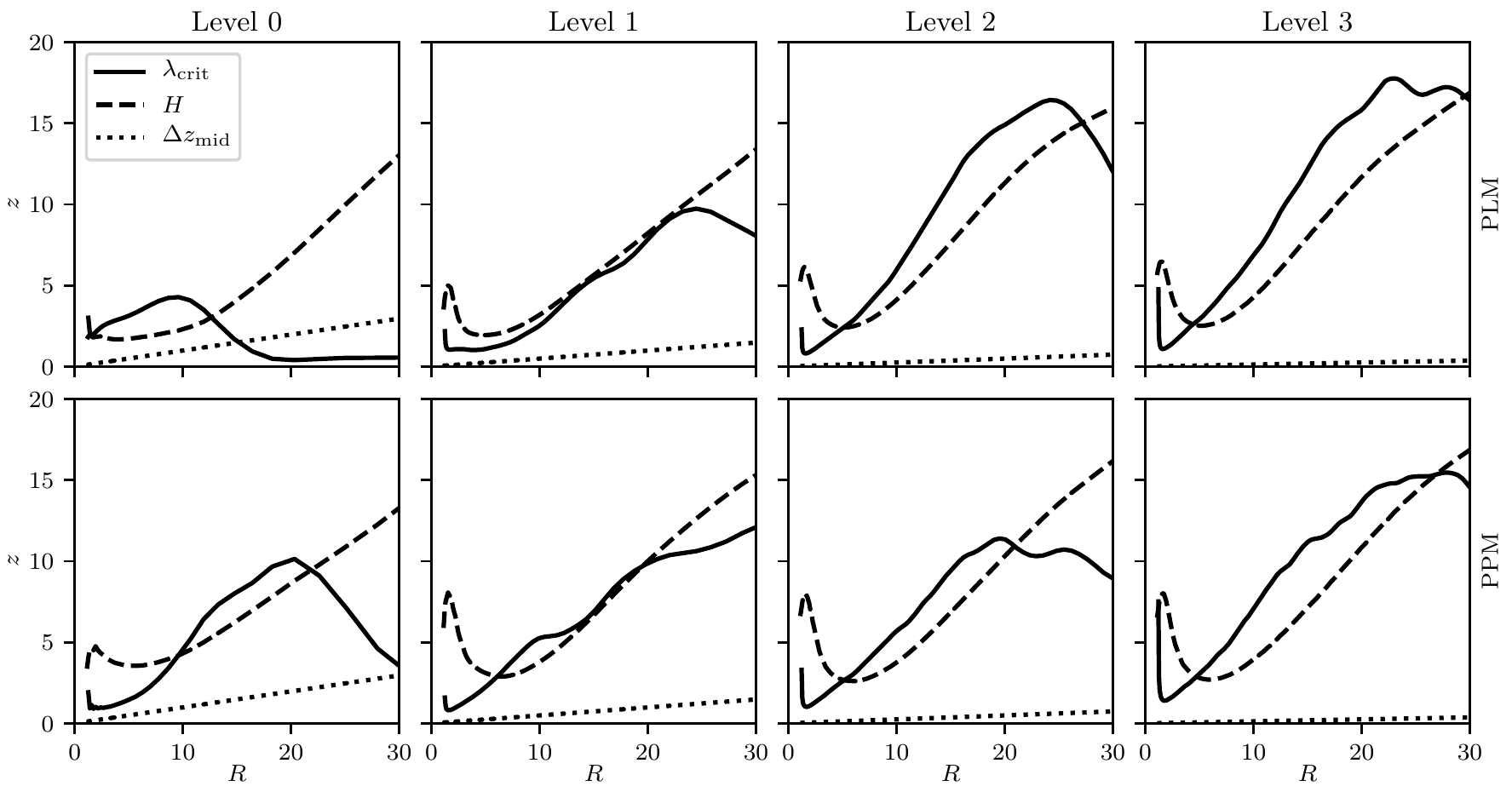}
  \caption{Radial profiles of MRI critical wavelength $\lambdac$ (solid), scale height $H$ (dashed), and midplane cell height $\Delta z\midp$ (dotted). Throughout much of the inner disk we have $\lambdac \geq H$, indicating the magnetic field has become strong enough to suppress the MRI. \label{fig:mri_wavelength}}
\end{figure*}

Figure~\ref{fig:mri_wavelength} shows the radial profiles of $H$ and $\lambdac$, each averaged in $\phi$ and $t$. Also shown is the height $\Delta z\midp$ of a single cell at the midplane. Except at the innermost few gravitational radii, $\lambdac$ reaches or exceeds the scale height throughout the inner regions of all the disks, indicating these are sufficiently MAD to inhibit the MRI. Disk scale height is relatively invariant across simulations, as is the run of $\lambdac$ in the regions that have reached steady state. At the lowest resolutions we are only marginally resolving $\lambdac$ and $H$, though at higher resolutions we have many cells per critical wavelength and per scale height.

We define an MRI suppression factor $S \equiv 2H/\lambdac$. This counts the number of critical wavelengths spanning the full disk, so the instability is expected to be unimportant for $S \lesssim 1$. Note the suppression factor in \citet{McKinney2012} differs from ours by a factor of $\fcrit/\sqrt{3}$, and so their cutoff of $1/2$ corresponds to $S \lesssim \sqrt{3}/(2\fcrit) \approx 1$ with our definition. The fact that $S$ is small even in the lowest resolutions, where we have already seen other evidence of underresolved behavior, indicates that building up enough vertical flux to suppress the MRI is only a matter of physical conditions and that any reasonable resolution is numerically capable of showing the effect.

\Citet{Gammie2004} also gives formulas corresponding to the maximum growth rate. This occurs at a wavelength of
\begin{equation}
  \lambdam \equiv 2\pi \vaz \sqrt{\frac{16 r^3}{15}} \fmax,
\end{equation}
where again we have a relativistic correction factor
\begin{equation}
  \fmax \equiv \frac{\sqrt{5} (1 + a r^{-3/2}) (1 - 3 r^{-1} + 2a r^{-3/2})^{1/2}}{(1 - 2 r^{-1} + a^2 r^{-2})^{1/2} (5 - 18 r^{-1} + 16 a r^{-3/2} - 3a^2 r^{-2})^{1/2}}
\end{equation}
multiplying the Newtonian version \citep[\extref{114} in][]{Balbus1998}. From this we can define a vertical MRI quality factor counting the number of cells in a single wavelength of the fastest growing mode:
\begin{equation}
  Q \equiv \frac{\lambdam}{r \Delta\theta}.
\end{equation}
Taking an $R$-weighted mean of $Q$ at the midplane inside $R = 10$, we find that the PLM simulations have average quality factors of $11$, $9$, $39$, and $95$, and the PPM simulations have values of $7$, $19$, $41$, and $99$. The lower two resolutions are only marginally resolving the MRI according to \citet{Hawley2011}.

% Disk properties: Disk structure
\subsection{Disk Structure}
\label{sec:disk:structure}

\begin{figure*}
  \centering
  \includegraphics{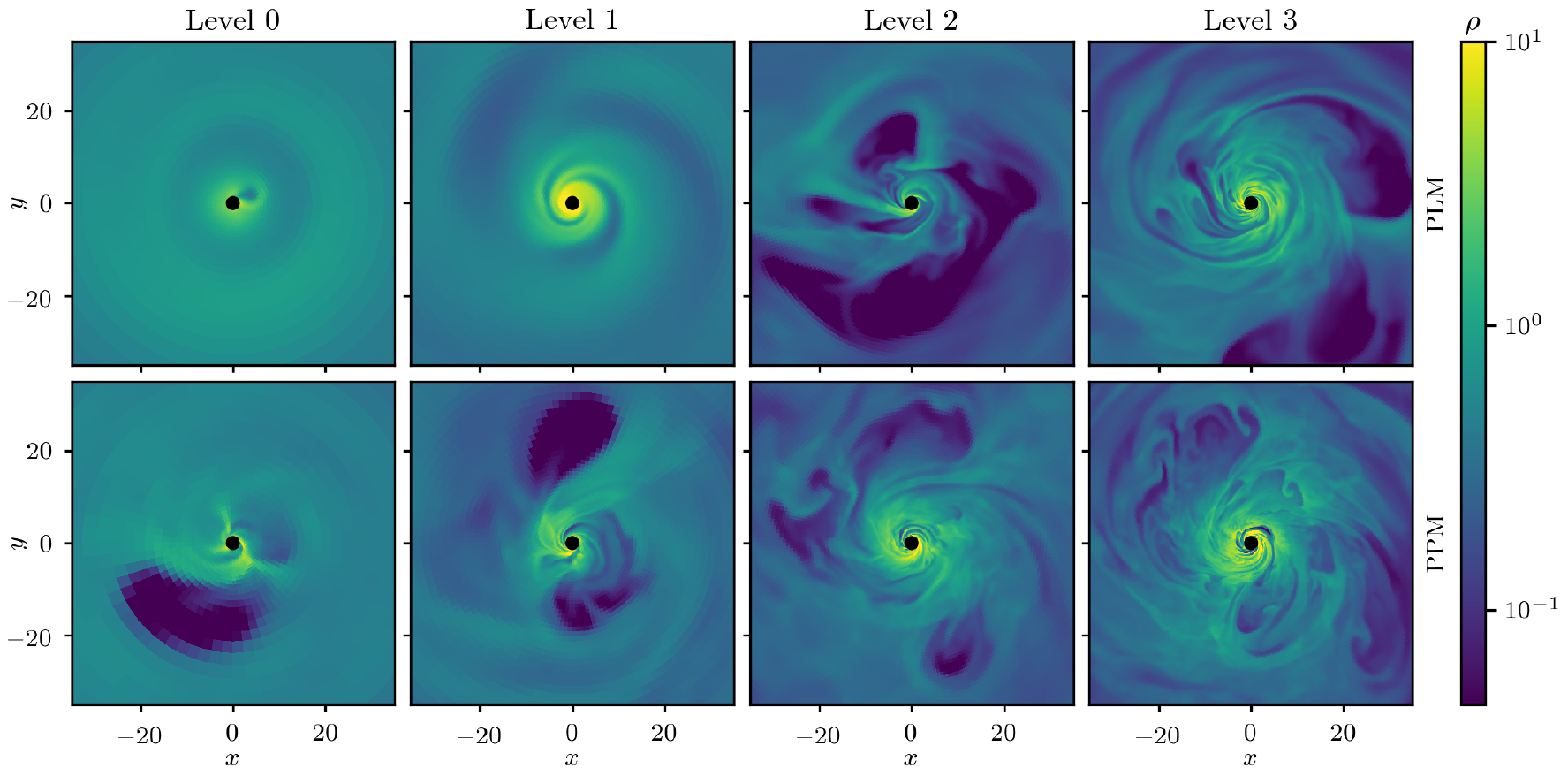}
  \caption{Density $\rho$ in the midplane at $t = 20{,}000$. Magnetic bubbles can be seen in all but the lowest resolutions, and their sizes are not dependent on resolution. However secondary instabilities make them less clearly defined with PPM at levels 2 and 3. \label{fig:slice_midplane_rho}}
\end{figure*}

Late-time snapshots of the midplane density structures of the disks for all our runs is shown in Figure~\ref{fig:slice_midplane_rho}. At the lowest resolutions little azimuthal variation is present. The remaining simulations show low-density magnetically supported bubbles repeatedly forming and rising buoyantly away from the black hole, as we expect from the magnetic RTI. These bubbles are well resolved where they appear, with their sizes being roughly independent of the grid scale. Still there is a clear trend of increasingly finer structure appearing at higher resolutions, with secondary instabilities appearing at the surfaces of the bubbles in the higher-resolution PPM simulations.

\begin{figure*}
  \centering
  \includegraphics{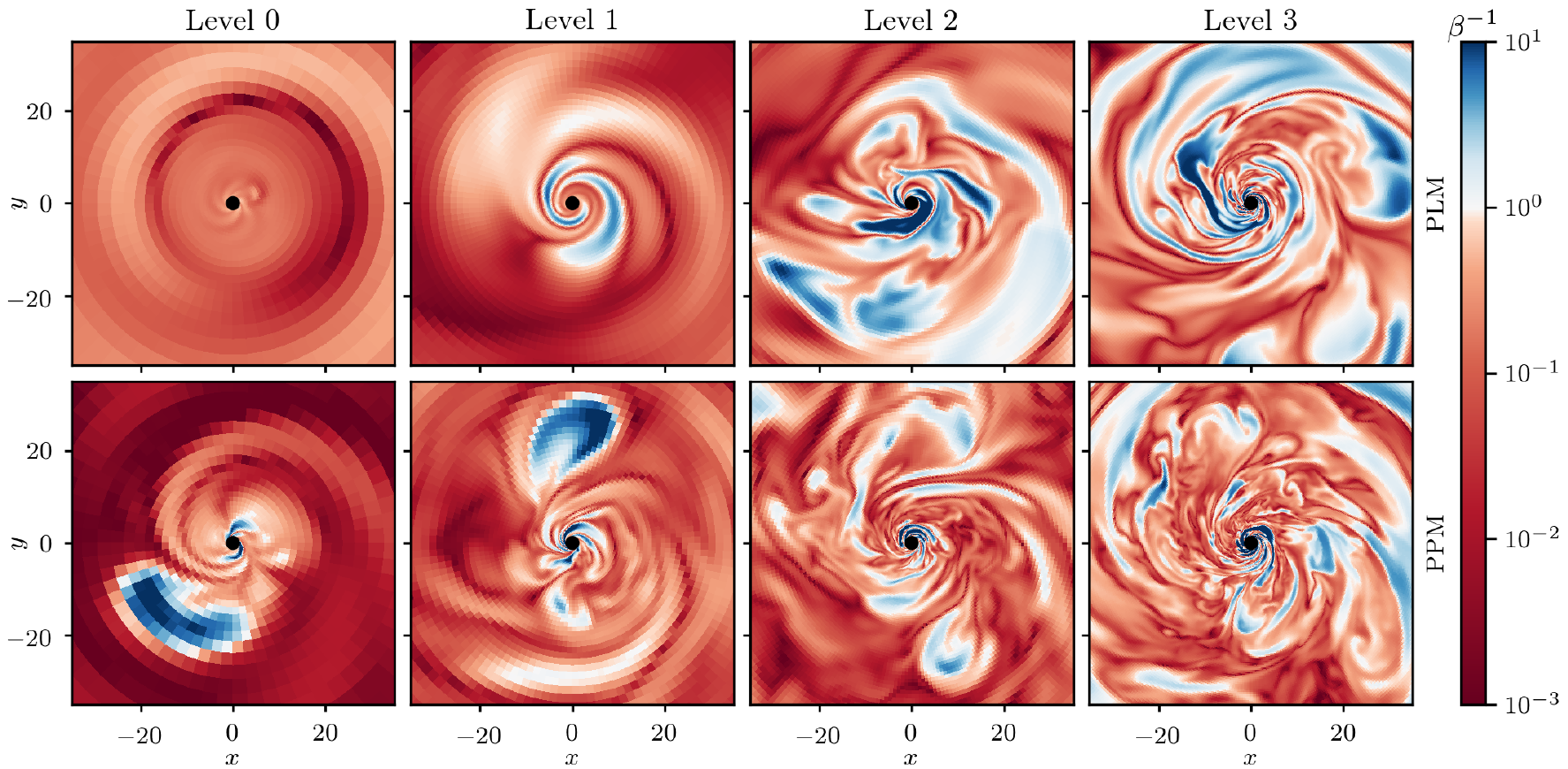}
  \caption{Plasma $\beta^{-1}$ in the midplane at $t = 20{,}000$. As resolution and reconstruction order increase, more and thinner current sheets appear and highly magnetized bubbles become more numerous. \label{fig:slice_midplane_beta_inv}}
\end{figure*}

Figure~\ref{fig:slice_midplane_beta_inv} shows plasma $\beta^{-1}$ in the same midplane slices. Dark red lines indicate field reversals and thus current sheets. They become thinner and more numerous at higher resolution and also at fixed resolution when transitioning from PLM to PPM.

\begin{figure*}
  \centering
  \includegraphics{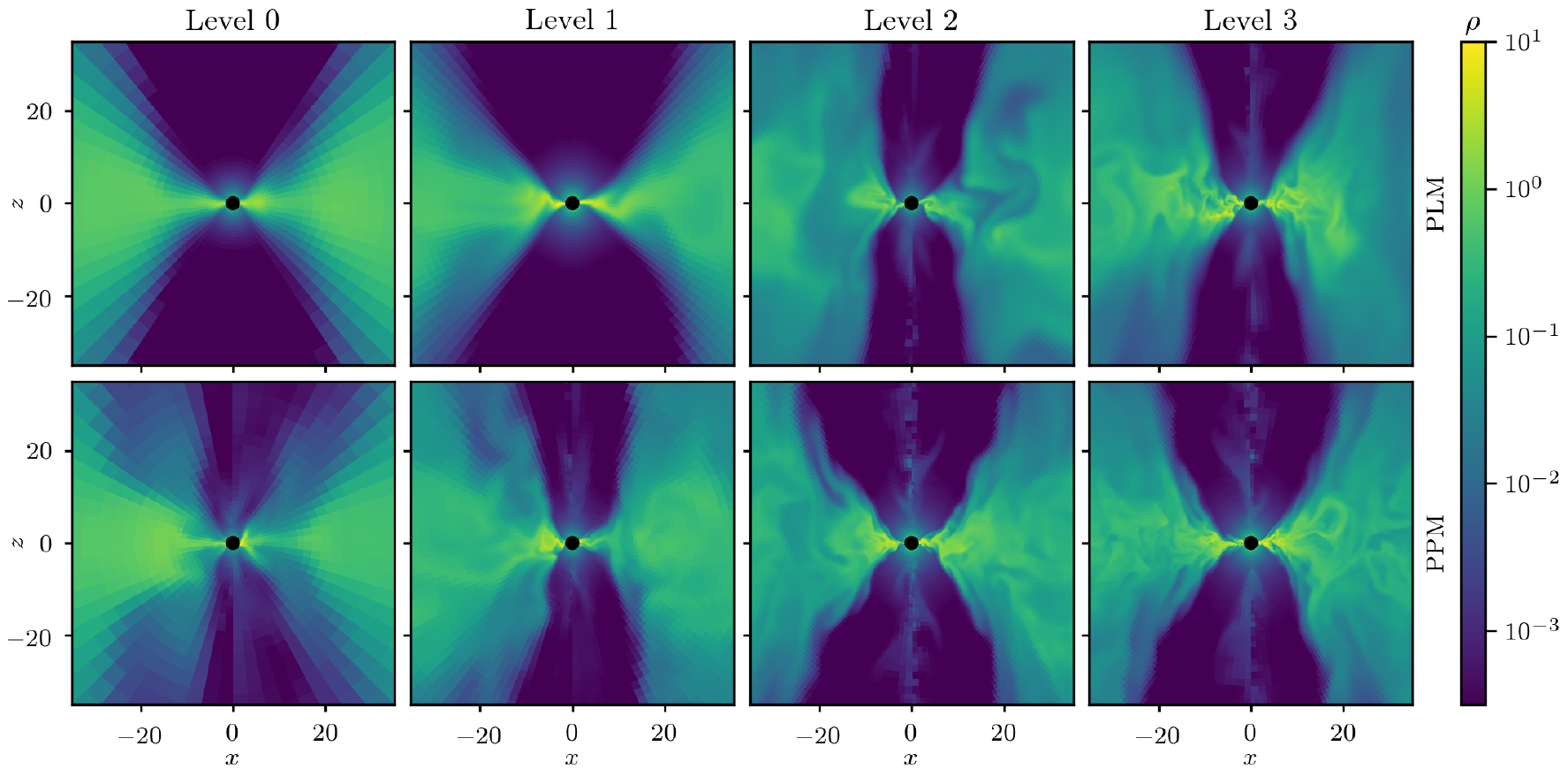}
  \caption{Density $\rho$ in a poloidal slice at $t = 20{,}000$. Turbulent structures appear in all but the lowest resolutions, but they are never resolved in the sense that they continue down to the grid scale in all cases. \label{fig:slice_polar_rho}}
\end{figure*}

\begin{figure*}
  \centering
  \includegraphics{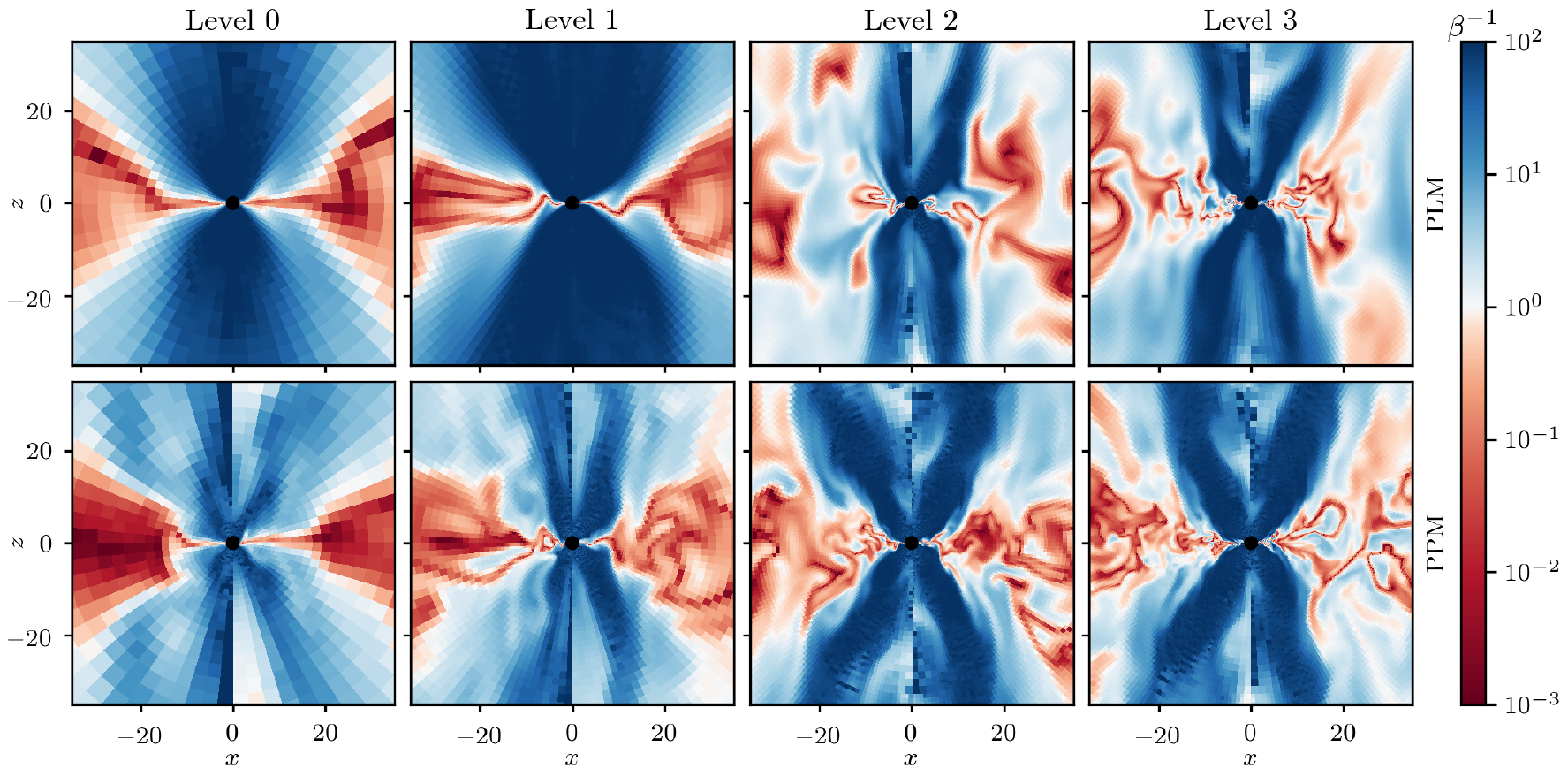}
  \caption{Plasma $\beta^{-1}$ in a poloidal slice at $t = 20{,}000$. The current sheet thicknesses are set by the grid scale. \label{fig:slice_polar_beta_inv}}
\end{figure*}

We also show late-time vertical slices through the simulation, colored by $\rho$ (Figure~\ref{fig:slice_polar_rho}) and $\beta^{-1}$ (Figure~\ref{fig:slice_polar_beta_inv}). In both cases there is structure down to the grid scale, with current sheets generally being a single cell thick.

The amount of structure can be more quantitatively expressed via correlation lengths of various quantities. We do this with correlations in the azimuthal direction in the midplane, following \citet{Shiokawa2012}. We define the time-dependent correlation function for a quantity $q$ to be
\begin{equation}
  R(t,r,\phi) = \int_0^{2\pi}\!\int_{\thetamin}^{\thetamax}\!\int_{\rmin}^{\rmax} \delta q(t, r', \theta', \phi') \delta q(t, r', \theta', \phi'+\phi) \, \dr'\,\dth'\,\dph',
\end{equation}
where $\delta q$ is the change in a quantity $q$ from its mean over the same domain, the range $\rmin < r < \rmax$ covers a single cell in radius, and the range $\thetamin < \theta < \thetamax$ covers the two cells in polar angle bordering the midplane. We normalize and average this to obtain the time-independent correlation function
\begin{equation}
  \bar{R}(r,\phi) = \frac{1}{\tmax - \tmin} \int_{\tmin}^{\tmax} \frac{R(t,r,\phi)}{R(t,r,0)} \, \dt.
\end{equation}

\begin{figure*}
  \centering
  \includegraphics{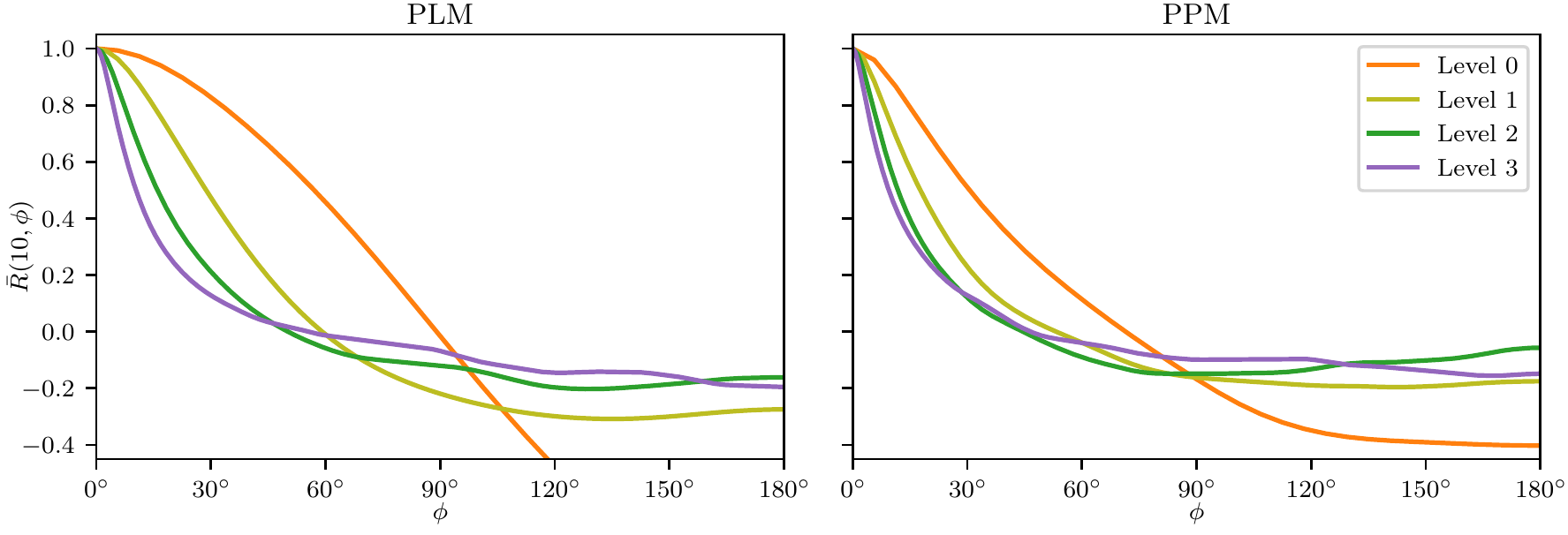}
  \caption{Time-independent correlation function for density $\rho$ at $r = 10$ as a function of azimuthal angle. There is some agreement at large scales. However, systematic trends are seen in small scales, with only the highest two resolutions using PPM close to agreeing. \label{fig:power_spectrum}}
\end{figure*}

Figure~\ref{fig:power_spectrum} shows the run of the time-independent correlation function with $\phi$ at representative value of $r = 10$, calculated for $q = \rho$. While large-scale features may be captured almost as well by the level~1 simulations as those at level 3, correlations at small scales are still changing with resolution between levels 2 and 3.

The essence of our correlation functions can be summarized by a single correlation length at each radius. The correlation length at a given radius $r$ is the value $\lambda$ for which $\bar{R}(r,\lambda) = \bar{R}(r,0)/\ee$. Note that these lengths are in fact angles.

\begin{figure*}
  \centering
  \includegraphics{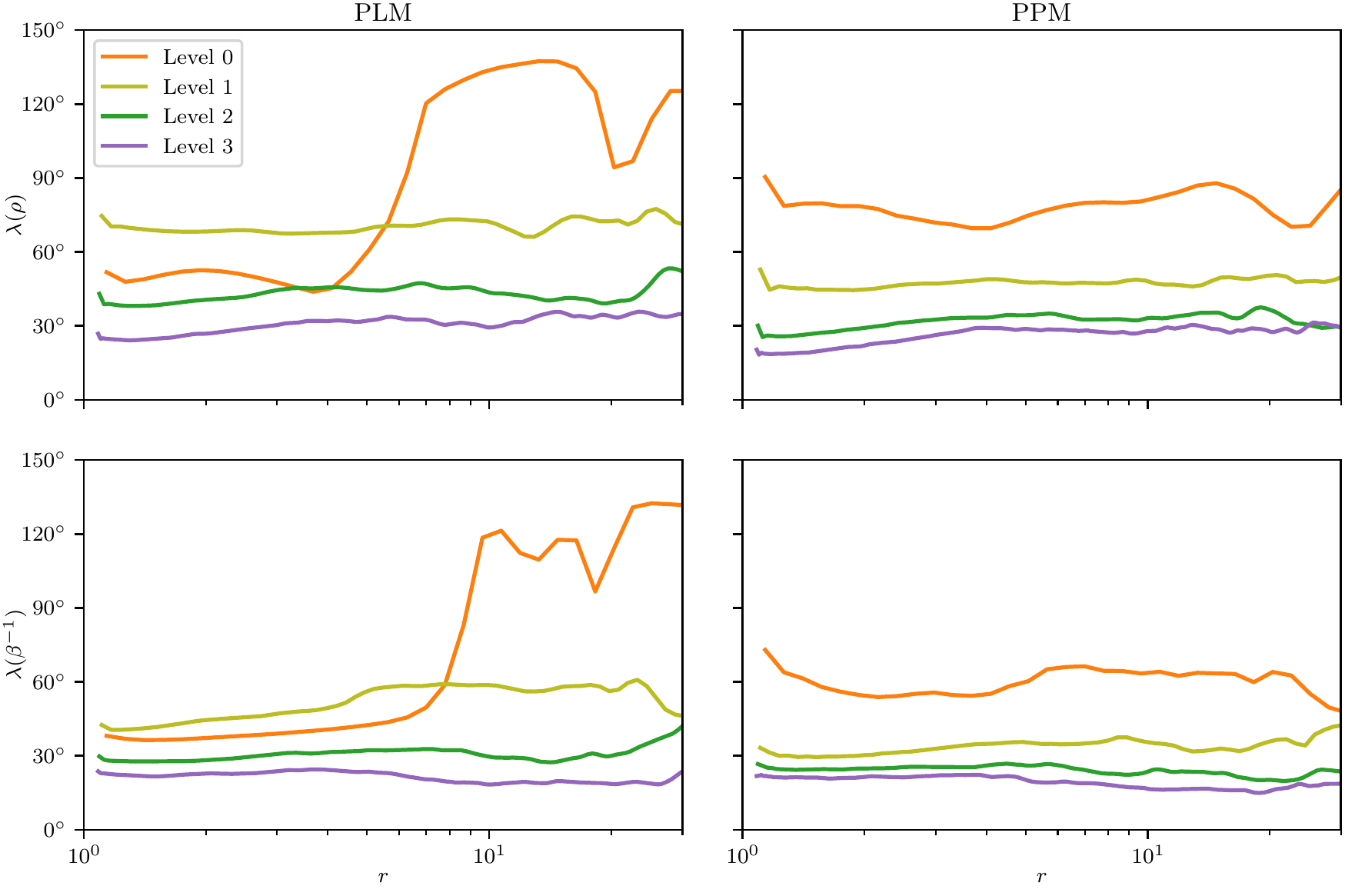}
  \caption{Midplane correlation length for $\rho$ (top) and $\beta^{-1}$ (bottom) in the azimuthal direction as a function of radius. Note the trend with resolution, indicating convergence has not been fully achieved. However, the line spacing indicates a trend toward convergence in the PPM cases, possibly achieved with another factor of $2\text{--}4$ in resolution. \label{fig:correlation_lengths}}
\end{figure*}

The correlation lengths for $\rho$ and $\beta^{-1}$ are shown in Figure~\ref{fig:correlation_lengths}. In all cases the angular length scales are invariant with radius over the inner region in which steady state has been achieved. At some fixed resolutions, correlation lengths in the plateau region are slightly lower when using PPM as opposed to PLM, and also slightly lower when looking at $\beta^{-1}$ rather than $\rho$, but these differences are small compared to the effect of resolution. We have not achieved convergence, though the values we measure suggest convergence might be achieved with approximately two more levels of refinement (a factor $4$ in resolution).

It is interesting to note that the correlation lengths shown in Figure~\ref{fig:correlation_lengths} are much larger than the azimuthal cell widths, which range from $5.625^\circ$ to $0.7031^\circ$. This is partly expected from having small structures sheared out in the azimuthal direction:\ the direction in which we measure correlations is not orthogonal to the current sheets. Furthermore, as noted in \citet{White2016} when comparing such correlations in non-MAD disks to those found by \citet{Shiokawa2012}, at fixed $\phi$-resolution midplane correlation lengths can decrease with increasing $\theta$-resolution in the midplane. That is, one should use caution in comparing these correlation lengths between simulations with different polar grids. We expect then that had we concentrated our grid near the midplane these particular correlation lengths would have been lower.

As there is no explicit dissipation mechanism in these simulations, it is expected that some turbulent properties continue to scale with the grid size and corresponding magnitude of numerical dissipation. However the non-convergence of small-scale turbulence does not necessarily preclude the convergence of global properties, especially in MAD flows. In contrast to cases with no net vertical flux, the transport of angular momentum in these flows can be dominated by large-scale magnetic stresses instead of stresses resulting from small-scale turbulence. This explains how our accretion rates can agree at sufficiently high resolution even when our correlation lengths do not.

% Jet properties
\section{Jet Properties}
\label{sec:jet}

% Jet properties: Magnetic structure
\subsection{Magnetic Structure}
\label{sec:jet:structure}

\begin{figure*}
  \centering
  \includegraphics{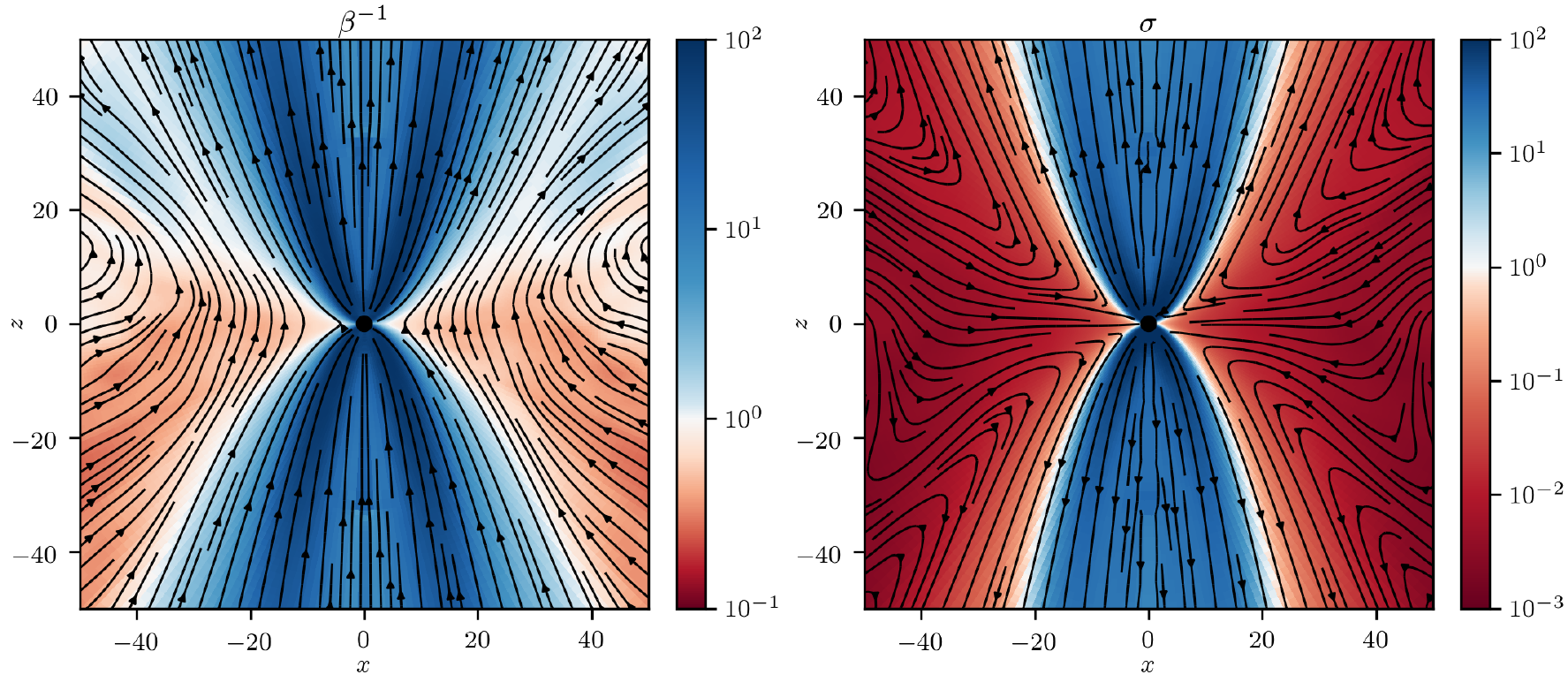}
  \caption{Plasma $\beta^{-1}$ (left) and $\sigma$ (right), averaged in azimuth (weighted by density) and time for the level~3 PPM run. The disk proper can be taken to be the red region of $\beta^{-1} < 1$ on the left, while the jet can be taken to be the blue region of $\sigma > 1$ on the right. The corona lies in between. Streamlines show the average poloidal part of the magnetic field (left) and density-weighted average of the poloidal part of the $4$-velocity (right). \label{fig:average_magnetization}}
\end{figure*}

While most of our simulations are quite turbulent, they do reach steady state in the average sense and so regions can be defined based on their $t$- and $\phi$-averaged properties. We do this with two measures of magnetization, plasma $\beta^{-1}$ and $\sigma$. Figure~\ref{fig:average_magnetization} shows the results for the level~3 PPM run (the other runs appear similar). As suggested in \citet{McKinney2012}, we can loosely define the disk to be where gas pressure dominates magnetic pressure, the jet to be where magnetic energy density dominates rest mass energy density, and the corona to be the region in between. The colorschemes in the figure emphasize these transitions at $\beta^{-1} \approx 1$ and $\sigma \approx 1$.

\begin{figure*}
  \centering
  \includegraphics{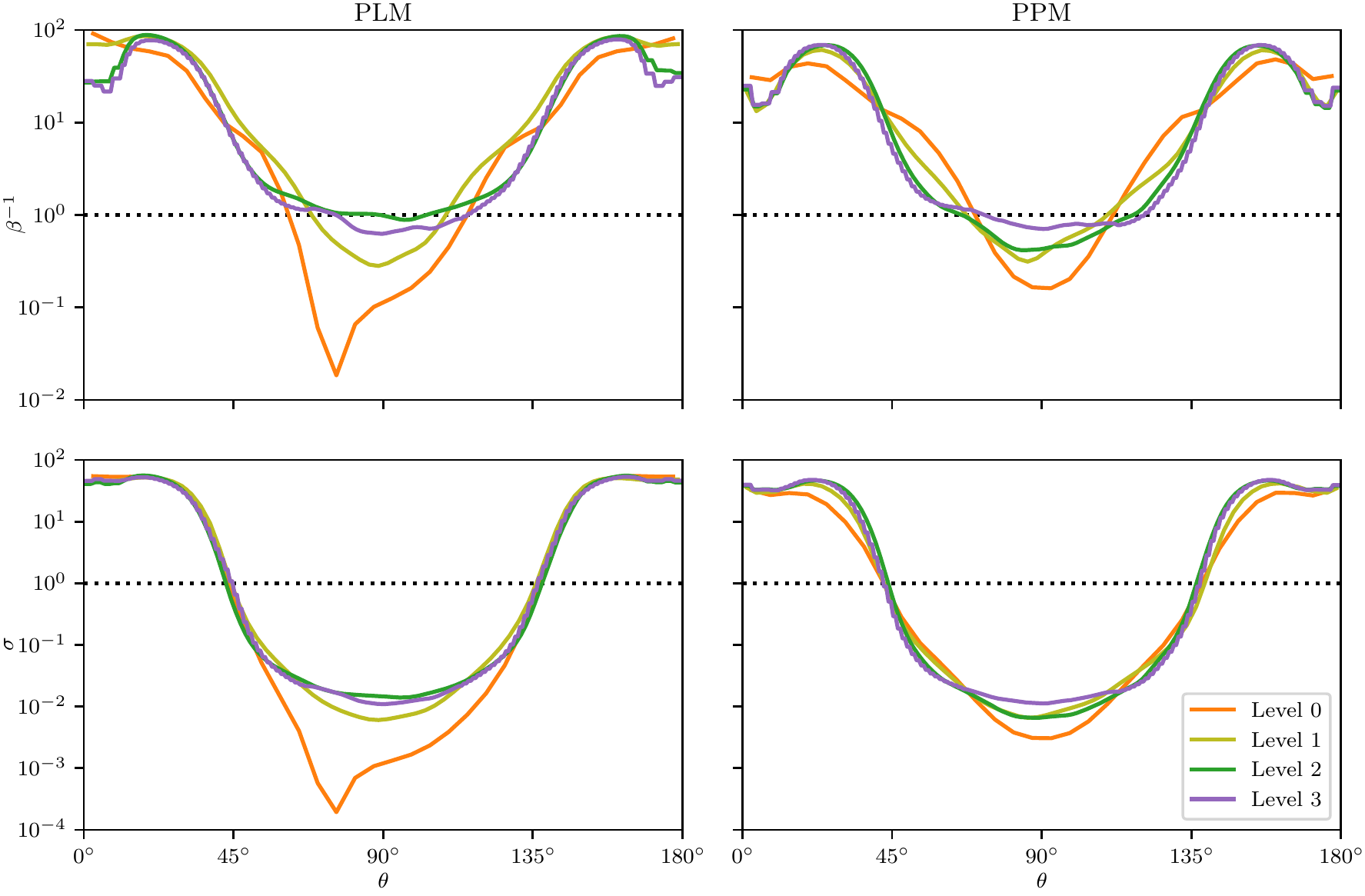}
  \caption{Polar angle profiles of plasma $\beta^{-1}$ (top) and $\sigma$ (bottom), averaged in azimuth and time. The values shown are at $r = 25$. There is good agreement in the jet region except at the lowest resolutions. At the same time, the middle of the disk continues to become more magnetized with increasing resolution, especially in terms of $\beta^{-1}$. At higher resolutions than those studied here, we may even expect $\pmag$ to exceed $\pgas$ everywhere. \label{fig:jet_magnetization}}
\end{figure*}

In order to compare our simulations, we sample average magnetization along arcs of constant $r = 25$ and varying $\theta$. The results are shown in Figure~\ref{fig:jet_magnetization}. The results across the simulations are in generally good agreement, with two exceptions. First, the lowest resolutions show slightly less magnetization in the bulk volume of the jet, outside the core. Second, there is a trend of disk magnetization increasing with resolution. While even very low resolutions can capture the qualitative structure of the magnetized accretion flow (agreeing with the findings in \S\ref{sec:disk:flux}), the quantitative amount of magnetization relative to the hydrodynamical state of the disk is another example of a property that has not entirely converged.

While $\beta^{-1}$ and $\sigma$ are large near the poles, they are below the numerical ceilings of $100$ at the sampled radius. We caution however that the magnetization could be larger closer to the horizon. Once the jet has propagated some distance it is no longer affected by the ceilings, but they could leave an imprint by adding density and pressure at very small radii.

% Jet properties: Energy extraction
\subsection{Energy Extraction}
\label{sec:jet:energy}

Just as with mass, we can calculate the flux of covariant (binding) energy through a spherical surface:
\begin{equation}
  \dot{E} \equiv \int\limits_{S_r} \tensor{T}{^r_t} \sqrt{-g} \, \dth\,\dph.
\end{equation}
The efficiency of the black hole is taken to be
\begin{equation}
  \eta \equiv \frac{\dot{M} - \dot{E}}{\ave{\dot{M}}},
\end{equation}
where the denominator is smoothed in time with a Gaussian filter of full-width at half-maximum $\Delta t = 100$. Given that $\ave{\dot{M}} > 0$ always (cf.\ Figure~\ref{fig:mdot_time}), $\eta > 0$ corresponds to some positive energy (negative binding energy) being transported to infinity. If this is simply the binding energy of a circular orbit at the ISCO relative to $0$ at infinity, as in the Novikov--Thorne model, we expect $\eta \approx 0.23$ for $a = 0.98$. In cases where $\eta > 1$, this energy exceeds not only the binding energy but also the rest-mass energy of the matter falling into the black hole, and we expect to see this given high spin and a MAD disk \citep{Tchekhovskoy2011}.

\begin{figure*}
  \centering
  \includegraphics{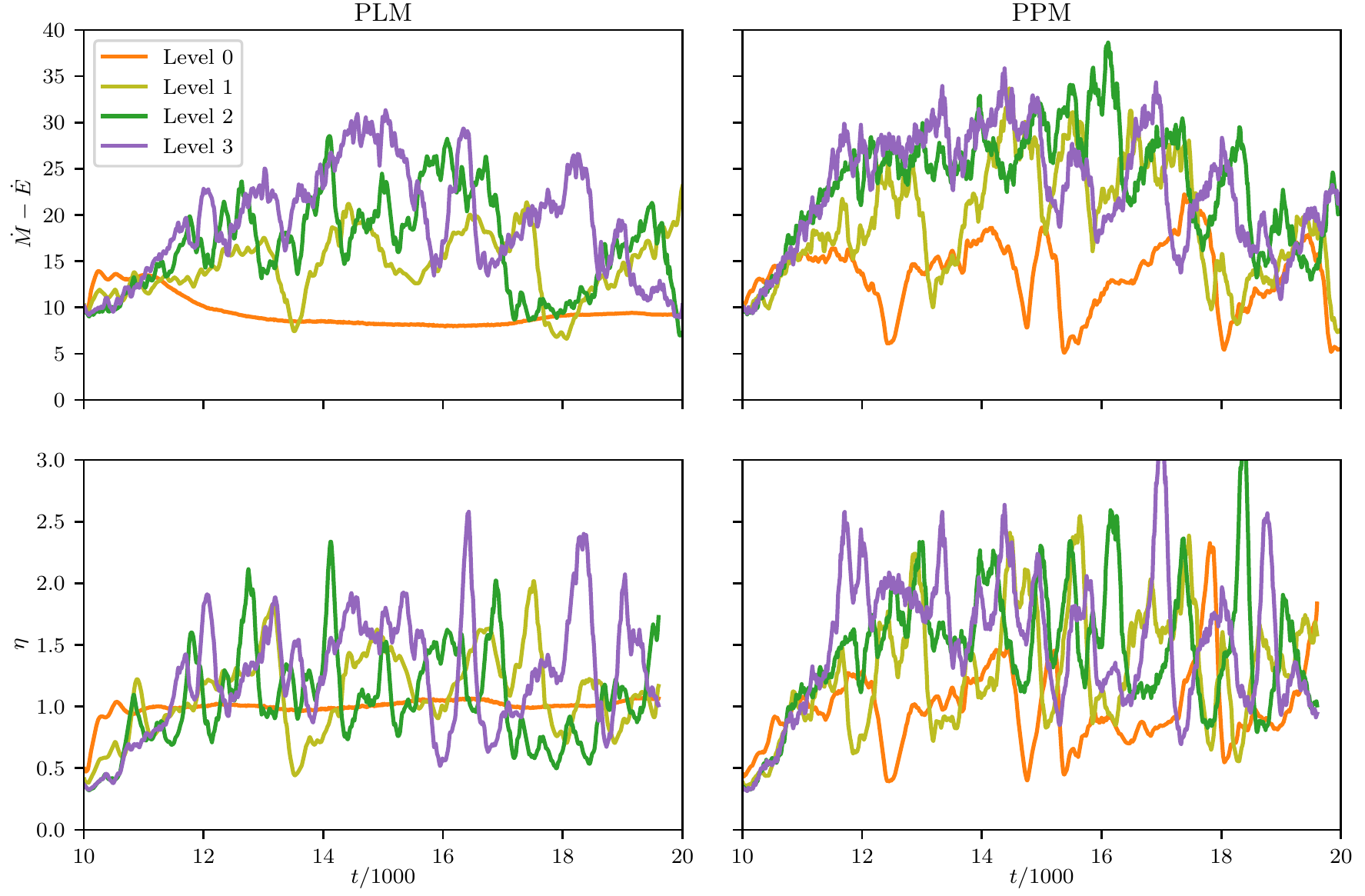}
  \caption{Flux of energy returned to infinity as a function of time (top), with the associated energy extraction efficiency (bottom). Large amounts of energy are returned, in most cases exceeding accreted rest-mass energy and yielding $\eta > 1$. \label{fig:edot_time}}
\end{figure*}

Figure~\ref{fig:edot_time} shows the energy extraction and efficiency as functions of time. As with other quantities that reflect the turbulent nature of these flows, the time series are highly variable, though again the level~0 PLM run shows no significant variability. At higher resolutions and with higher-order reconstruction, the energies and efficiencies are slightly higher, though the differences are somewhat obscured by the high variability.

Most of the measured efficiencies exceed unity, going up to approximately $150\%$. Even the level~0 PLM run shows $\eta \approx 1$, whereas we would expect $\eta \approx 0$ if the accretion flow were not in the MAD state, or if the black hole were not spinning.

We can differentiate between energy in the jet and energy in the wind by labeling locations as one of ``jet,'' ``wind,'' or ``disk'' according to the average accretion state as pictured in Figure~\ref{fig:average_magnetization}. Constructing these averages for all eight simulations, we define the jet to be where $\sigma > 1$, the wind to be where $\sigma \leq 1$ and $u^r > 0$, and the disk to be the remaining volume. The disk contributes negligibly (less than $1\%$) in all cases to the energy flux used to calculate $\eta$. Among the PLM simulations, the jet comprises $76\%$, $78\%$, $73\%$, and $73\%$ of the energy flux in order of increasing resolution. For PPM, these values are $86\%$, $78\%$, $79\%$, and $75\%$.

The jet contributions to $\eta$ alone are far larger than the Novikov--Thorne efficiency. Thus the Blandford--Znajek mechanism for extracting spin energy from the black hole is robustly modeled, even at moderate resolutions that do not fully capture other aspects of the accretion flow.

% Jet properties: Jet structure
\subsection{Jet Structure}
\label{sec:jet:outflow}

\begin{figure*}
  \centering
  \includegraphics{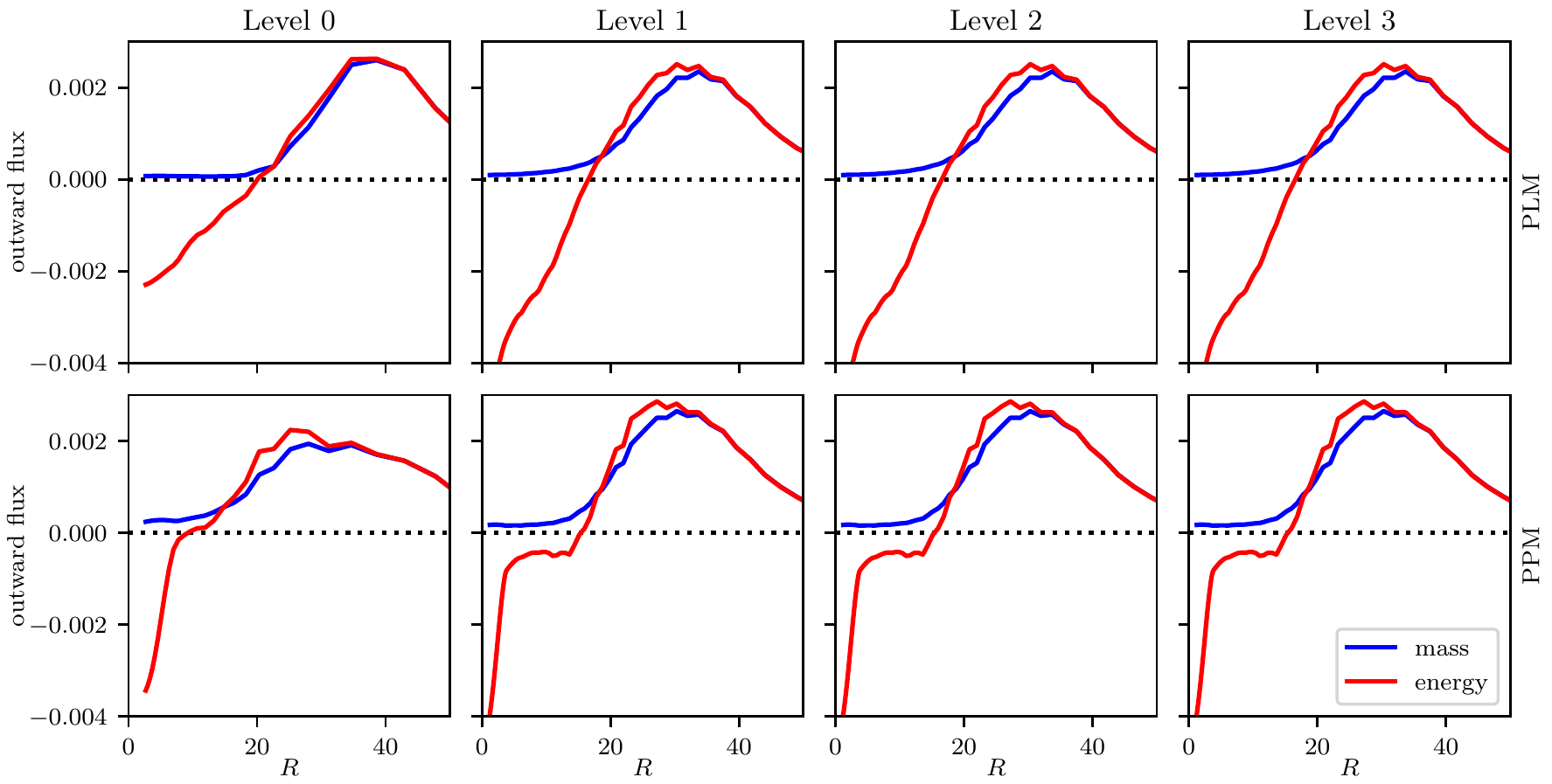}
  \caption{Fluxes of mass and binding energy per unit area ($\pm\rho u^z$ and $\pm\tensor{T}{^z_t}$) through the planes $z = \pm50\ M$, averaged in azimuth and time. Outward Poynting flux dominates the jet core. Changing the resolution does little to alter these profiles, though using PPM instead of PLM reconstruction helps to resolve the jet. \label{fig:jet_flux}}
\end{figure*}

While the rate of spin energy extraction may not vary much with resolution, the jet structure can change. Figure~\ref{fig:jet_flux} shows cylindrical radial profiles of fluxes in the jet in the $z = \pm50$ planes (averaging the results for the two planes). In particular, we plot the outward mass flux per unit area $\pm\rho u^z$ and outward binding energy per unit area $\pm\tensor{T}{^z_t}$. This later quantity is negative where energy is flowing to infinity. Total fluxes through the planes correspond to integrating the values against $\sqrt{-g\cyl} \, \drr\,\dvph$.

In these planes, the jet ($\sigma \gtrsim 1$) is inside $R \approx 20$, as can be seen in Figure~\ref{fig:jet_magnetization}. Immediately outside the jet there is considerable coronal outflow of mass in all simulations. Inside the jet region, the mass flux becomes negligible, with a large amount of energy flowing outward in the form of Poynting flux.

Our refinement scheme refines the jet in these planes, though the jet base is always on a coarse grid. Any resolution-dependent effects we see are therefore due to processes in the jet but beyond the base, or else due to properties of the disk. In fact there is very little variation in jet flux profiles with disk resolution. However, all PPM simulations consistently concentrate much of the energy flux in a narrow jet core, while the PLM simulations have a much broader distribution in $R$. This suggests the jet structure would change if the resolution of the always-coarse jet base were altered, likely appearing more like the PPM results at higher resolution.

\begin{figure*}
  \centering
  \includegraphics{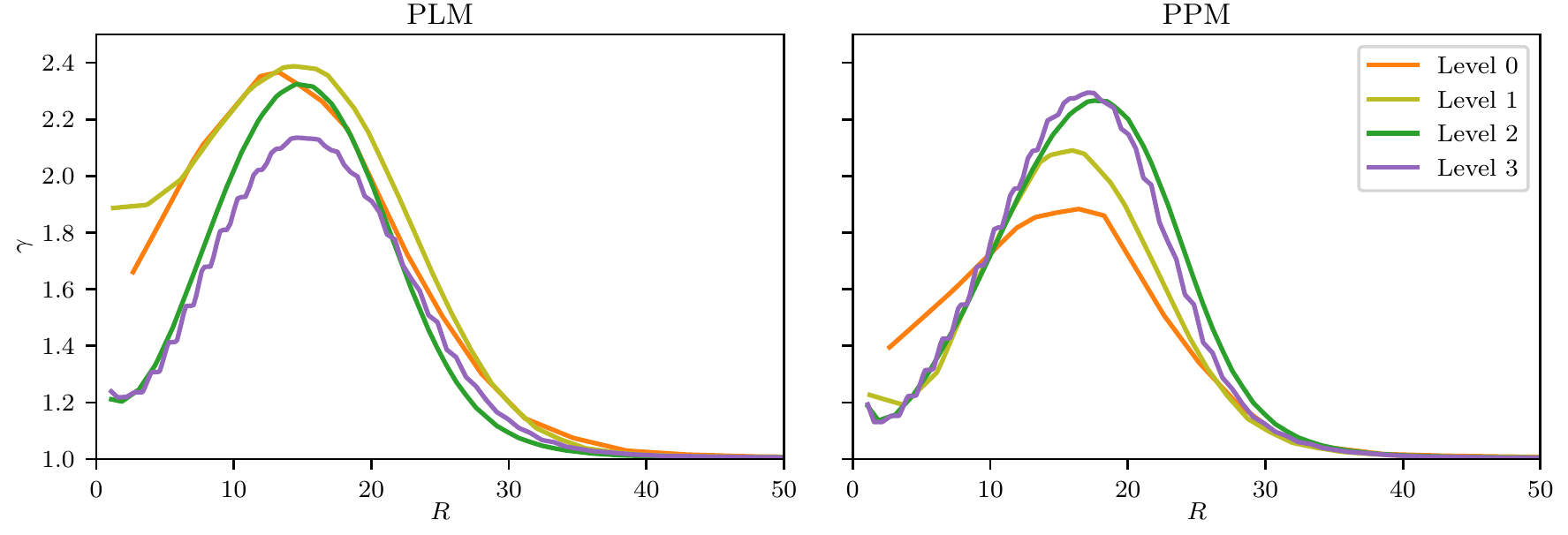}
  \caption{Normal-frame Lorentz factor $\gamma$ of the fluid in the $z = \pm50$ planes, averaged in azimuth and time. There is considerable variation with resolution, with only a hint of convergence in the level~2 and level~3 PPM runs. All profiles show the cores of the jets to be slower than the walls. The step-like nature of the level~3 profiles is an artifact of prolongating coarser data to refinement level 3 before analysis. \label{fig:jet_gamma}}
\end{figure*}

Figure~\ref{fig:jet_gamma} shows the Lorentz factor of the fluid, relative to the normal observer, through these same planes. Here the effects of resolution are more noticeable. The similarity of the level~2 and level~3 PPM profiles might indicate convergence, though the grid resolution in the jet region does not change between these two refinement schemes. Qualitatively, all runs show similar off-axis peaks in Lorentz factor, with the core being slower than the jet walls.

We caution that properties of this highly magnetized, rapidly moving part of the flow can depend numerically on not only resolution, but also the treatment of floors and variable inversion failures. Radial profiles of mass and internal energy fluxes indicate that density and gas pressure floors add only very small amounts of material and heat in our simulations. Still, the ideal MHD condition means that even with perfect magnetic field evolution, electric fields and thus electromagnetic energies are sensitive to any changes in fluid velocity resulting from numerical floors.

% Light curves
\section{Light Curves}
\label{sec:light_curves}

The results of the previous sections show some of the physical properties of the accretion flow are converged with resolution while others are not. This raises the question of how converged are direct observables. While our simulations do not model photons in situ, we can postprocess snapshots in order to generate light curves from the simulations. For this purpose we choose to use the code \ibothros{}, which models synchrotron emission and absorption and propagates photons along geodesics in the Kerr spacetime to generate an image of specific intensity at a given frequency.

We must set several physical scales, and we choose these to reflect the supermassive black hole in M87. Following \citet{Ryan2018} we set the black hole mass to $M = 6.2\times10^{9}\ \msun$, placed at a distance of $D = 16.7\ \mpc$. We choose a viewing angle $20^\circ$ off the jet axis. The electron density is taken to trace the fluid density, with a constant temperature ratio of $4$ between the ion and electron temperatures, and the electrons are assumed to be thermal. Images are sampled every $10\ GM/c^3$ (i.e.\ every $3.5$ days) over the simulations times $15{,}000 < t / (GM/c^3) \leq 20{,}000$ (i.e.\ spanning $4.8$ years). Each image consists of $200^2$ pixels covering a patch of sky $30\ GM/c^2D$ ($110\ \mmas$) on each side. The density scale of the fluid is adjusted until the total emission at $230\ \ghz$ matches the observed value of $0.98\pm0.04\ \jy$ from \citet{Doeleman2012}.

\begin{figure}
  \centering
  \includegraphics{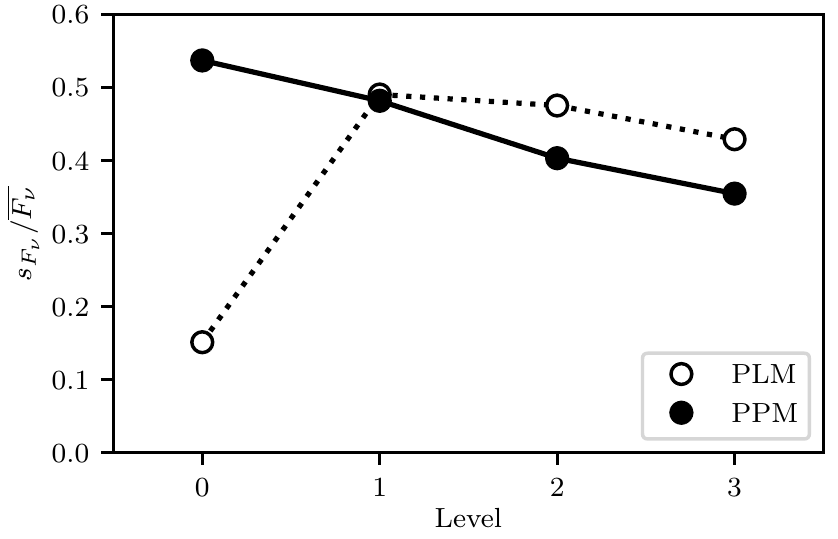}
  \caption{Light curve variability for the eight simulations. As with other properties, the level~0 PLM simulation does not show signs of turbulence, but in all other cases the light curves reach a saturated state and display large variations about that mean. The variability is still decreasing with resolution even at level 3. \label{fig:variability}}
\end{figure}

Except in the level~0 PLM case, where the flux varies monotonically with time, the light curves all display variability about a saturated state. The standard deviations of the fluxes are shown in Figure~\ref{fig:variability}. The standard deviations of these standard deviations are small, less than $0.02$ in all cases, making the trend with resolution statistically significant. That is, variability is decreasing with increasing resolution, and it does not appear converged at our highest resolution.

% Discussion
\section{Discussion}
\label{sec:discussion}

In examining the same problem of MAD accretion at different resolutions, we can determine which properties have converged and are therefore the most robustly determined by simulations.

The large-scale structure of the accretion flow is relatively constant across our simulations. In particular, the time-averaged magnetization as depicted in Figure~\ref{fig:average_magnetization} roughly holds at all resolutions. This is reflected in the constancy of where the solid lines cross the dotted lines in the lower panels of Figure~\ref{fig:jet_magnetization}, as well as the similar sizes of magnetic bubbles shown in Figures~\ref{fig:slice_midplane_rho} and~\ref{fig:slice_midplane_beta_inv}.

Figure~\ref{fig:mri_wavelength} shows that the MRI is robustly suppressed in the inner part of all our simulations. Even at low resolutions, we can see that the critical wavelength of the MRI tracks the disk scale height. Our initial setup thus robustly leads to a MAD state in the sense of growth of the MRI being suppressed by buildup of magnetic flux and compression of the disk. This occurs even at lower resolutions where the MRI quality factors indicate we are only moderately sampling unstable wavelengths.

A number of important global parameters converge at sufficiently high resolution, including the accretion rate (Figures~\ref{fig:mdot_radius} and~\ref{fig:mdot_time}), the vertical magnetic flux near the horizon (Figure~\ref{fig:flux_radius} and~\ref{fig:flux_time}), and the rate of energy extraction from the black hole (Figure~\ref{fig:edot_time}). This last quantity is highly variable in time in the MAD state, and so longer run times would allow for a more precise measurement of the mean efficiency.

We consistently find about $3$ times as much outgoing energy in the magnetized jet as in the wind. This is in agreement with the values of $\eta_\mathrm{j}$ and $\eta_\mathrm{w,o}$ reported in Table~5 of \citet{McKinney2012} for comparable models (A0.9N25, A0.9N50, A0.9N100, A0.9N200, and A0.99N100). On the other hand, the MAD models of \citet{Avara2016} and \citet{MoralesTeixeira2018} have more energy leaving through the wind than the jet. This can be an effect of the thinness of those radiatively cooled disks, as well as the fact that they only have $a = 0.5$, meaning the Blandford--Znajek process will produce a much weaker jet.

Other aspects of the flow show signs of converging, though there is disagreement at our highest resolutions, indicating somewhat higher resolutions are required to attain convergence. For example, the midplane values of magnetization as measured by $\sigma$ agree at high resolutions, but there is still disagreement when using $\beta^{-1}$ (Figure~\ref{fig:jet_magnetization}). The quantitative measure of MRI suppression, $S$, follows $\beta^{-1}$ in not being fully converged.

The statistical nature of the turbulence, as measured by correlation lengths, has also not quite converged (Figures~\ref{fig:power_spectrum} and~\ref{fig:correlation_lengths}); size scales decrease as resolution increases (and they decrease slightly as higher-order reconstruction is used). Figure~\ref{fig:power_spectrum} in particular is essentially a power spectrum and captures the same information as the azimuthal power spectra computed by \citet{MoralesTeixeira2018}, who see agreement among simulations that only have up to $128\times128\times64$ cells within $r \approx 10$. They have at most $7$ azimuthal cells per correlation length and see convergence while we have up to $43$ cells per correlation length ($512$ cells with a length of $30^\circ$) and do not have the same agreement. This discrepancy is likely due to the radiative nature of their disks.

It may seem inconsistent that we easily suppress the MRI even at low resolutions and yet we do not attain convergence in turbulence at our highest resolution. This tension is resolved by having another driving force for turbulence, the magnetic RTI. \Citet{Marshall2018} study the stresses in a thinner MAD disk, and they also see turbulent angular momentum transport despite MRI suppression, attributing this to the constant formation and disruption of Rayleigh--Taylor bubbles.

The jet Lorentz factor (Figure~\ref{fig:jet_gamma}) is another quantity whose values may still change with higher resolution. It is somewhat surprising that energy extraction shows less discrepancy between different resolutions than Lorentz factor. We reiterate that the jet base has the same grid in all our simulations, so we can only measure convergence in terms of properties that depend on resolving disk properties and resolving the propagation of the jet tens of gravitational radii beyond the horizon.

Some properties are far from converged in our simulations. The thinnest current sheets are always at the grid scale (Figures~\ref{fig:slice_midplane_beta_inv} and~\ref{fig:slice_polar_beta_inv}), and secondary instabilities on top of Rayleigh--Taylor bubbles do not even appear except at the highest resolutions (Figure~\ref{fig:slice_midplane_rho}).

More disconcerting is the trend light curve variability has with resolution (Figure~\ref{fig:variability}). We are possibly far from convergence, which one expects if the short-timescale variability comes from small, transient, highly magnetized parcels of plasma that are not well resolved on the grid.

We summarize key average quantities for the simulations in Table~\ref{tab:summary}. For the accretion rate $\dot{M}$ and horizon magnetic flux $\varphi$ we take a simple average in time, quoting the standard deviation of the $501$ samples in the range $15{,}000 \leq t \leq 20{,}000$. We do the same for the efficiency $\eta$ but with $40$ fewer samples, as times at the very end of time simulation cannot be used due to the width of the Gaussian filter used in calculating these quantities. For the MRI suppression factor $S \equiv 2H/\lambdac$, we average the scale height $H$ and critical wavelength $\lambdac$ in azimuth, calculate $S$, and then average in time. Then $S$ and the density correlation length $\lambda(\rho)$ have been time averaged, but we must find a way to average in radius. For $q \in \{S, \lambda(\rho)\}$ we define
\begin{equation}
  \bar{q} = \frac{\int qR\,\drr}{\int R\,\drr}
\end{equation}
and
\begin{equation}
  \sigma^2_q = \frac{\int q^2R\,\drr}{\int R\,\drr} - \bar{q}^2,
\end{equation}
where the integrals are over the region $R < 10$, using $20$, $40$, $79$, and $159$ samples at levels 0 through 3. The table reports $\bar{q} \pm \sigma_q$ for all five of these measures. For the $N = 500$ values of $F_\nu$ in each light curve, we define $s^2$ to be the unbiased estimator of the variance of the set, with the uncertainty in $s$ taken to be
\begin{equation}
  \sigma_s = s \paren[\bigg]{1 - \frac{2}{N-1} \paren[\bigg]{\frac{\Gamma(N/2)}{\Gamma((N-1)/2)}}^2}^{1/2}
\end{equation}
in terms of the gamma function. The table reports $(s \pm \sigma_s) / \overline{F_\nu}$, where the mean used to normalize is always within $0.01\ \jy$ of the observed $0.98\ \jy$.

\floattable
\begin{deluxetable}{cD@{ $\pm$}DD@{ $\pm$}DD@{ $\pm$}DD@{ $\pm$}DD@{ $\pm$}DD@{ $\pm$}D}
  \tablecaption{Summary of simulation results. \label{tab:summary}}
  \tablewidth{0pt}
  \tablehead{\colhead{Run} & \multicolumn4c{$\dot{M}$} & \multicolumn4c{$\varphi$} & \multicolumn4c{$S$} & \multicolumn4c{$\lambda(\rho)$} & \multicolumn4c{$\eta$ (\%)} & \multicolumn4c{$s_{F_\nu}/\overline{F_\nu}$}}
  \decimals
  \startdata
  PLM--0 &  8.39 & 0.59 & 51.2 & 1.0 & 0.190 & 0.029 & 1.76  & 0.60  & 103.1 &  2.5 & 0.1513 & 0.0048 \\
  PLM--1 & 12.8  & 3.8  & 53.2 & 7.1 & 0.69  & 0.18  & 1.243 & 0.034 & 118   & 29   & 0.490  & 0.016  \\
  PLM--2 & 15.7  & 6.2  & 51.1 & 8.0 & 0.47  & 0.47  & 0.788 & 0.026 & 107   & 36   & 0.475  & 0.015  \\
  PLM--3 & 15.8  & 7.2  & 52.7 & 8.8 & 0.37  & 0.34  & 0.545 & 0.027 & 133   & 48   & 0.429  & 0.014  \\
  PPM--0 & 13.5  & 4.3  & 40.9 & 8.8 & 0.96  & 0.34  & 1.360 & 0.060 & 100   & 37   & 0.537  & 0.017  \\
  PPM--1 & 14.9  & 6.0  & 52.1 & 8.6 & 0.56  & 0.57  & 0.834 & 0.014 & 141   & 43   & 0.482  & 0.015  \\
  PPM--2 & 16.3  & 6.7  & 57.3 & 7.6 & 0.47  & 0.47  & 0.575 & 0.023 & 153   & 51   & 0.403  & 0.013  \\
  PPM--3 & 15.6  & 6.5  & 56.4 & 7.5 & 0.37  & 0.33  & 0.480 & 0.027 & 147   & 52   & 0.355  & 0.011  \\
  \enddata
\end{deluxetable}

Of the quantities in the table, the horizon-penetrating flux $\varphi$ is the most clearly converged, showing similar values even at the lowest resolutions. The accretion rate $\dot{M}$, MRI suppression factor $S$, and efficiency $\eta$ also appear converged at the resolutions of refinement levels 2 and 3, though these are intrinsically highly variable statistics in these flows. Turbulent correlation length $\lambda(\rho)$ and synchrotron light curve variability $s_{F_\nu}/\overline{F_\nu}$ do not converge by level 3.

% Conclusion
\section{Conclusion}
\label{sec:conclusion}

We study a single physical scenario of MAD accretion onto a rapidly spinning black hole at various resolutions. This is done employing the static mesh refinement ability of \athena, spanning a factor of $8$ in resolution in each dimension over the bulk of the disk. By only varying resolution and comparing results, we can determine which features of such simulations are adequately resolved and can therefore be trusted. A priori, it is not obvious how much sensitivity to resolution we should expect in MAD structure given inward field advection is competing with the complex nonlinear processes of turbulent diffusion and magnetic RTI.

Certain global properties are found to converge with resolution in our set of simulations, notably accretion rate $\dot{M}$ and jet efficiency $\eta$, as well as the fact that the MRI is suppressed. Magnetic structure is generally well captured. In particular, the accumulated magnetic flux on the horizon $\varphi$ is found to be essentially the same at all resolutions we consider. Buoyant magnetic bubbles are well resolved on the higher resolution grids (though with secondary Kelvin--Helmholtz instabilities only appearing at high resolution with higher-order spatial reconstruction).

As simulations in the literature use grids that go up to our levels 2 and 3 \citep{Tchekhovskoy2011,McKinney2012,Avara2016}, going even higher in some cases \citep{Liska2018}, these results provide assurance that the large-scale structures of MAD flows are being accurately modeled by the community.

On the other hand, the details of jet structure, including profiles of energy flux and Lorentz factor, do not agree among our simulations. Higher resolutions at the base of the jet itself would be helpful in reliably determining the velocity and energy distributions between the core and outer parts of the jet. The Lorentz factor too will probably change at higher resolutions, though increasing resolution alone will very likely not be enough to create jets with $\gamma \sim 10\text{--}100$ given that $\gamma$ in all our jets peaks between $1.8$ and $2.4$.

We do not see convergence in the length scales of turbulent eddies, though these scales may converge with another factor of approximately $4$ in resolution. We also do not see convergence in the thicknesses of current sheets, and indeed this is not surprising given the lack of any explicit dissipation mechanism decoupled from the grid scale in the simulations.

The lack of convergence in small-scale features would not necessarily be a concern, except these features can affect observables. One might hope that in going from accretion flows with disordered magnetic fields to MAD flows, with small-scale turbulence giving way to large-scale magnetic stresses as the important transport mechanism, the emitted light will be less sensitive to the resolution. However, when postprocessing our simulations to model thermal synchrotron emission, we see that there is light curve variability that continues to scale with resolution even at the highest resolutions (Figure~\ref{fig:variability}). There are certainly caveats to these findings --- including not modeling the radiation dynamically, not including other radiative processes such as Compton scattering, and not considering nonthermal particles --- indicating future avenues of exploration. Still, our results suggest caution in comparing simulations to data via light curves, even when most aspects of the simulation appear converged.

We conclude that the general MAD flow structure is robust. This includes the suppression of the MRI as well as the field structure that results from an equilibrium between inward advection and the rising of magnetically buoyant bubbles in a setting unstable to RTI. Details of turbulence are however not fully captured at these resolutions, in the sense that some observable quantities are affected by our choice of grid, and this can affect observables.

We relegate the computationally expensive study of resolution at the base of the jet to a separate investigation. Future simulations with in situ radiative transfer can help further determine to what extent the variation of turbulence with resolution has on direct observables.

% Acknowledgments
\acknowledgments

We thank S.~M. Ressler for help in using \ibothros{} on our data. This work used the Extreme Science and Engineering Discovery Environment (XSEDE) Stampede2 at the Texas Advanced Computing Center through allocation AST170012, resources of the Argonne Leadership Computing Facility (ALCF) via the Innovative and Novel Computational Impact on Theory and Experiment (INCITE) program, and the Savio computational cluster resource provided by the Berkeley Research Computing program at the University of California, Berkeley. This work was supported in part by a Simons Investigator award from the Simons Foundation and NSF grants AST~13-33612 and AST~17-15054 (EQ).

% Software
\software{\athena{} \citep{White2016}, \ibothros}

% References
\bibliographystyle{aasjournal}
\bibliography{references}

\end{document}